# Tailoring Interlayer Charge Transfer Dynamics in 2D Perovskites with Electroactive Spacer Molecules

Yorrick Boeije,[1,2†] Wouter T.M. Van Gompel,[3†] Youcheng Zhang,[2,4†] Pratyush Ghosh,[2] Szymon J. Zelewski,[1,2] Arthur Maufort,[3] Bart Roose,[1] Zher Ying Ooi,[1] Rituparno Chowdhury,[2] Ilan Devroey,[3] Stijn Lenaers,[3] Alasdair Tew,[2] Linjie Dai,[1,2] Krishanu Dey,[2] Hayden Salway,[1] Richard H. Friend,[2] Henning Sirringhaus,[2] Laurence Lutsen,[3] Dirk Vanderzande,[3] Akshay Rao,[2*] Samuel D. Stranks[1,2*]

AUTHOR ADDRESS [1]Department of Chemical Engineering and Biotechnology, University of Cambridge, Philippa Fawcett Drive, Cambridge, CB3 0AS, UK.

[2]Department of Physics, Cavendish Laboratory, University of Cambridge, JJ Thomson Ave, Cambridge, CB3 0HE, UK.
[3]Hasselt University, Institute for Materials Research (IMO-IMOMEC), Hybrid Materials Design (HyMaD), Martelarenlaan 42, B-3500 Hasselt, Belgium.

[4]Cambridge Graphene Centre, Department of Engineering, University of Cambridge, JJ Thomson Ave, Cambridge, CB3 0FA, UK.

**ABSTRACT:** The family of hybrid organic-inorganic lead-halide perovskites are the subject of intense interest for optoelectronic applications, from light-emitting diodes to photovoltaics to X-ray detectors. Due to the inert nature of most organic molecules, the inorganic sublattice generally dominates the electronic structure and therefore optoelectronic properties of perovskites. Here, we use optically and electronically active carbazole-based Cz-$C_i$ molecules, where $C_i$ indicates an alkylammonium chain and $i$ indicates the number of $CH_2$ units in the chain, varying from 3–5, as cations in the 2D perovskite structure. By investigating the photophysics and charge transport characteristics of $(Cz-C_i)_2PbI_4$, we demonstrate a tunable electronic coupling between the inorganic lead-halide and organic layers. The strongest interlayer electronic coupling was found for $(Cz-C_3)_2PbI_4$, where photothermal deflection spectroscopy results remarkably demonstrate an organic-inorganic charge transfer state. Ultrafast transient absorption spectroscopy measurements demonstrate ultrafast hole transfer from the photoexcited lead-halide layer to the Cz-$C_i$ molecules, the efficiency of which increases by varying the chain length from $i$=5 to $i$=3. The charge transfer results in long-lived carriers (10–100 ns) and quenched emission, in stark contrast with the fast (sub-ns) and efficient radiative decay of bound excitons in the more conventional 2D perovskite $(PEA)_2PbI_4$, in which phenylethylammonium (PEA) acts as an inert spacer. Electrical charge transport measurements further support enhanced interlayer coupling, showing increased out-of-plane carrier mobility from $i$=5 to $i$=3. This study paves the way for the rational design of 2D perovskites with combined inorganic-organic electronic properties through the wide range of functionalities available in the world of organics.

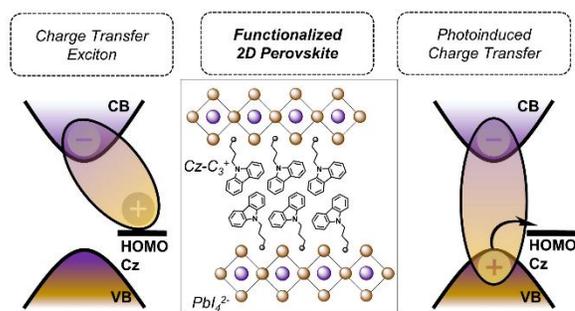

## INTRODUCTION

Hybrid inorganic-organic metal-halide perovskites, having a formula of $ABX_3$ in which A is a monovalent organic cation, B a divalent metal cation, and X a halide anion, have gained significant interest for their application in optoelectronic devices, such as photovoltaics, owing to their high absorption coefficients, long diffusion lengths, and high charge carrier mobilities.[1–4] Similarly, layered (2D) perovskites (formula of $A_2BX_4$) have shown enormous promise for various optoelectronic applications, including light-emitting diodes, field-effect transistors and photovoltaics.[5,6] 2D perovskites traditionally incorporate large, electronically insulating A-cations in the crystal structure (Figure 1a), resulting in strongly bound electron-hole pairs (i.e. excitons) which are localized within the 2D $PbI_4^{2-}$ layers due to dielectric and quantum confinement effects.[7–9] Consequently, the electronic coupling between the inorganic lead-halide layers is weak, with the electronically insulating organic cations creating a large tunneling barrier. Additionally, these organic cations only have an indirect effect on the electronic structure by causing structural distortions of the inorganic layer and/or by modifying the distance between the inorganic layers. As such, they do not directly contribute to any functionality of 2D perovskites.

To expand the functionalities of 2D perovskites, one could envision enhancing the electronic interactions between the inorganic and organic layers by incorporating π-conjugated organic cations with frontier orbitals close to the band edge states of the inorganic layers (Figure 1b and 1c). Density functional theory calculations on theoretically designed Dion-Jacobson 2D perovskites have indicated that electronically active (or "electroactive") organic spacers increase the out-of-plane electronic coupling through enhanced orbital overlap and may result in interlayer excitons through the formation of inorganic-organic hybridized orbitals.[10] The electroactive ligand could also result in a type II band alignment, which is in contrast to the type I band alignment for conventional 2D perovskites due to the very wide bandgap of the organic species that are generally used (Figure 1a and 1b); there, organic frontier orbitals are energetically far removed from the inorganic band edge states resulting in fully inorganic-dominated exciton physics.[11–13] Tailoring exciton dynamics,[14] energy-transfer induced phosphorescence,[15–19] charge-transfer induced photoluminescence quenching,[20–23] and exciton binding energy engineering[24] are some of the most recent demonstrations of 2D perovskites functionalized with electronically active organic molecules. Moreover, the addition of bulky π-conjugated ligands has resulted in stable and efficient perovskite optoelectronic devices.[25–33] Vertical (i.e. perpendicular to the $PbI_4^{2-}$ layers) charge transport in functionalized 2D perovskites has been predicted computationally[34,35] and was demonstrated experimentally in 2D perovskite crystals containing pyrene and perylene organic cations.[36] Vertical charge transport is an important additional functionality relevant for optoelectronic applications, such as photovoltaics, as 2D $PbI_4^{2-}$ layers usually grow parallel to the substrate surface and perpendicular to the out of plane direction of charge transport required within (photovoltaic) devices.[37]

Pushing these devices towards new functionality and performance will require an absolute control over optical properties, charge/exciton recombination and transport.[38,39] Previous time-resolved studies on functionalized 2D perovskites have reported interlayer photoinduced charge transfer (CT), although direct spectral observation of the CT state has proven difficult due to overlapping spectral signatures.[21,40–43] For instance, Gélvez-Rueda et al. performed microwave conductivity and transient absorption (TA) measurements to show the presence of long-lived charges in the inorganic layer formed faster than 200 fs after selective photoexcitation of the organic layer composed of a charge transfer complex.[42,44] A joint transient reflection-photoluminescence spectroscopic study on thiophene-based 2D perovskites characterized the CT process with a lifetime of 10 ps, and a long-lived species in the 600-700 nm range was assigned to the CT state, although the nature of this species was not further elucidated.[43]

Currently, interlayer photoinduced CT in 2D perovskites has not been studied in a controlled manner through systematic variation of the inorganic-organic interlayer distance. Moreover, despite the above-mentioned computational prediction of interlayer CT excitons, the optically active behavior of the spacer molecule – next to its electroactive behavior – through observation of an interlayer CT transition has not been explored.

Here, we investigate the interlayer inorganic-organic charge transfer dynamics of thin films of a series of carbazole-alkylammonium based 2D perovskites, $(Cz-C_i)_2PbI_4$, in which $i$ varies from 3 to 5 and contrast them to the conventional $(PEA)_2PbI_4$ system. We provide the first evidence of a direct optical interlayer charge transfer transition in a 2D perovskite using a combination of photothermal deflection spectroscopy and TA spectroscopy with sub-15 fs temporal resolution. TA spectroscopy is an ideal tool as it is capable of not only temporally tracking the CT processes as demonstrated in the above-mentioned reports, but also by directly probing the characteristic photoinduced absorption of the electronically distinct radical species generated after charge transfer.[45] In this case the $(Cz-C_i)^{\cdot+}$ radical cation (i.e. a hole polaron localized on $Cz-C_i$) could be detected. By modifying the alkyl chain length from $C_5$ to $C_3$, we observe a rise in the CT quantum yield by decreasing the organic-inorganic distance. The charge transfer-dominated excited state dynamics in $(Cz-C_i)_2PbI_4$ is consistent with a spatial separation of the electron and hole wavefunctions, resulting in a nanosecond carrier lifetime. We ascribe this to a reduced radiative probability, in contrast to $(PEA)_2PbI_4$ where the electron-hole pair remains strongly bound in the inorganic layer and decays rapidly with a time constant of 74 ps. Finally, the enhanced electronic interaction between the inorganic and organic layers in $(Cz-C_i)_2PbI_4$ is established by measuring electrical vertical charge transport using the space charge limited current (SCLC) method. These results establish direct observations of charge transfer and transport in 2D perovskite systems with electroactive molecular cations, paving the way for a large library of materials with new optoelectronic function.

RESULTS AND DISCUSSION

**Steady-state Characterization**

We first discuss the material characterization and optical properties of $(PEA)_2PbI_4$ and $(Cz-C_i)_2PbI_4$. The syntheses of the $Cz-C_i^+$ cations as well as the 2D perovskites are described in the experimental section and the supporting information. Schematic structures of $(PEA)_2PbI_4$ and $(Cz-C_4)_2PbI_4$ are shown in Figure 1a and Figure 1c, respectively. Figure 1b shows the proposed qualitative band alignments in both materials, where $(PEA)_2PbI_4$ has a type I alignment and $(Cz-C_i)_2PbI_4$ type II. The latter was proposed in ref.[23], considering that the valence band maximum (VBM) of lead-iodide 2D perovskites ($\sim$-6.0 eV)[46,47] is typically energetically deeper than HOMO-Cz (-5.4 eV),[48,49] and the conduction band minimum (CBM) ($\sim$-3.6 eV)[46,47] is far deeper than LUMO-Cz (-1.74 eV).[49] Smooth films with thicknesses of $\sim$17 nm, 38 nm, 34 nm and 46 nm were obtained for $(PEA)_2PbI_4$,

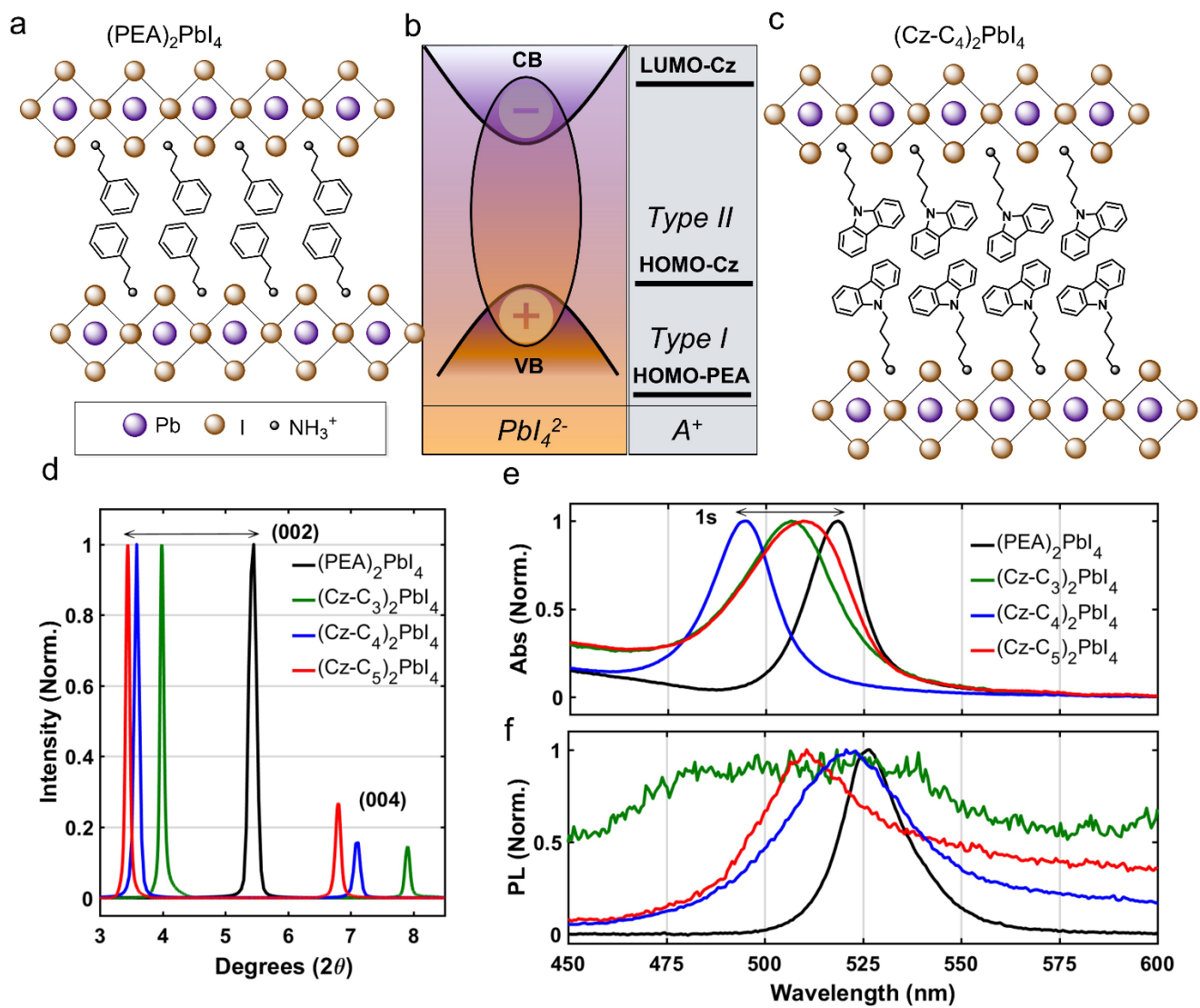

**Figure 1.** (a) Schematic structure of (PEA)$_2$PbI$_4$. Purple (orange) circles indicate Pb (iodide) atoms. (b) Schematic band diagram for (Cz-C$_4$)$_2$PbI$_4$ and (PEA)$_2$PbI$_4$, showing the conduction band (CB) and the valence band (VB) of the inorganic sublattice (assumed to be the same for both materials), as well as the highest occupied molecular orbitals (HOMO) for both organic fragments. The lowest unoccupied molecular orbital (LUMO) is only shown for Cz. The type II band alignment in (Cz-C$_4$)$_2$PbI$_4$ makes the organic cation electro-active, whereas the type I band alignment of (PEA)$_2$PbI$_4$ corresponds to an electro-inactive organic cation. (c) Schematic structure of (Cz-C$_4$)$_2$PbI$_4$. Herein, Cz-C$_4$ indicates carbazole-butylammonium. (d) X-ray diffraction patterns of (Cz-C$_n$)$_2$PbI$_4$ and (PEA)$_2$PbI$_4$ thin films. Only the first two reflections are shown for (Cz-C$_n$)$_2$PbI$_4$ and the first reflection for (PEA)$_2$PbI$_4$. (e) UV/VIS Absorption, normalized at the excitonic peak (1s). (f) Normalized photoluminescence (PL) spectra. $\lambda_{exc}$ = 400 nm.

(Cz-C$_3$)$_2$PbI$_4$, (Cz-C$_4$)$_2$PbI$_4$ and (Cz-C$_5$)$_2$PbI$_4$, respectively, by spin-coating precursor solutions of PbI$_2$ and PEAI (or Cz-C$_i$I) (Figure S1 and Experimental Section for solution processing details). Figure 1d shows XRD patterns for films of (PEA)$_2$PbI$_4$ and (Cz-C$_i$)$_2$PbI$_4$ (see Figure S2 for the patterns over a wider range of angles). We identify the equally spaced reflections that are characteristic for thin films of 2D perovskites.[50,51] From this, we determine the interplanar spacing (d-spacing) (note that we use the term "interplanar" to distinguish between the PbI$_4^{2-}$-Cz "interlayer" distance). As expected, the lead-halide interplanar reflections corresponding to a spacing of 22.3Å, 25.1Å and 26.0Å for (Cz-C$_3$)$_2$PbI$_4$, (Cz-C$_4$)$_2$PbI$_4$ and (Cz-C$_5$)$_2$PbI$_4$, respectively, are larger than that of (PEA)$_2$PbI$_4$ (16.3Å) due to the longer alkyl tail and bulkier aromatic moiety in the A-cation. Note that the increase in d-spacing from $i$=3 to $i$=5 is not linear, potentially due to different penetrations depths of the ammonium cation into the inorganic lattice.[36] In absorption spectra of the thin films, we observe that the first bright excitonic transition (1s) in (Cz-C$_4$)$_2$PbI$_4$ is blue-shifted by ~21 nm (~0.1 eV) with respect to the 1s peak in (PEA)$_2$PbI$_4$ (Figure 1e). Such an observation could be the result of i) strong octahedral tilting in the PbI$_4^{2-}$ layers due to the sterically demanding Cz-C$_4$ cation resulting in a larger band-gap,[52–54] and

ii) the higher dielectric constant of the organic layer which could result in a smaller exciton binding energy.[55–57] Both effects i) and ii) might also rationalize the broader 1s transition in (Cz-C$_i$)$_2$PbI$_4$, which could be further enhanced by strong exciton-phonon coupling in the case of a CT exciton via the Fröhlich interaction.[58–60] Investigating the origin of this blueshift, as well as the decreased blueshifts in (Cz-C$_3$)$_2$PbI$_4$ and (Cz-C$_5$)$_2$PbI$_4$ will be the subject of future work. Another absorption peak appears at 349 nm for (Cz-C$_i$)$_2$PbI$_4$ (Figure S3) which corresponds to the first excited state of the Cz-C$_i$ molecules as demonstrated by the absorption spectrum of the control (Cz-C$_i$)I films (Figure S4).

The PL spectra in the 450-600 nm range of (PEA)$_2$PbI$_4$ and (Cz-C$_i$)$_2$PbI$_4$ are shown in Figure 1f. Whereas a sharp strong excitonic emission peak characteristic for typical 2D perovskites appears at 525 nm for (PEA)$_2$PbI$_4$,[61] the emission spectra for the (Cz-C$_i$)$_2$PbI$_4$ films are characterized by much weaker and broader peaks. The complete (unnormalized) spectra of (Cz-C$_i$)$_2$PbI$_4$ (Figure S5) also show broad red emission, which has been frequently observed in 2D perovskites and is either explained by the presence of deep defects[62–64] or self-trapped excitons.[65,66] Although the red emission feature is a subject of further study, we do confirm that it is not present in emission spectrum of the (Cz-C$_i$)I salt (Figure S6) and is therefore not related to emission from the organic species. The significantly weaker and broader PL in (Cz-C$_i$)$_2$PbI$_4$ compared to (PEA)$_2$PbI$_4$ serves as the first observation of the impact of the electroactive spacer on the electronic properties of the 2D perovskite.

**Sub-gap Charge Transfer State**

We further investigated the optical properties of the sub-gap region by performing photothermal deflection spectroscopy (PDS) measurements, as shown in Figure 2a. A broad sub-gap state at ~600 nm is revealed for (Cz-C$_3$)$_2$PbI$_4$ (green line), which is indicative of a charge transfer (CT) state as commonly observed in organic photovoltaic blends.[67–69] The external quantum efficiency (EQE) spectrum of a (Cz-C$_3$)$_2$PbI$_4$ device made in a solar cell stack (see Figure S20 and Table S1 for J-V curves and corresponding parameters, respectively) replicates this feature, which appears to be more efficient at photocurrent extraction than the 1s exciton considering the ~100x smaller absorption coefficient only results in a drop in EQE of ~10x. CT states are typically associated with weak and broad PL spectra, as also observed for (Cz-C$_3$)$_2$PbI$_4$ (Figure 1f), due to poor orbital overlap between the spatially separated electron and hole wavefunctions. A sub-gap CT state would be consistent with the type II band alignment presented in Figure 1b and implies a non-zero oscillator strength associated with the transitions from the HOMO-Cz to the PbI$_4^{2-}$ CB. However, this feature is absent in both the PDS and EQE spectra of (PEA)$_2$PbI$_4$ (black curves in Figure 2a and 2b, respectively) as expected for a type I band alignment associated with the electro-inactive organic spacer.

To validate that the sub-gap feature is associated with an interlayer CT state, we measure the sub-15 fs ultrafast transient absorption spectra of (Cz-C$_3$)$_2$PbI$_4$ after photoexcitation with a broad-band pulse centered at 600 nm (see red spectrum, Figure 2a). A photoinduced absorption (i.e. negative ΔT/T) band at ~800 nm appears (blue spectrum) and is assigned to the Cz radical cation, from now on referred to as Cz$^{+\cdot}$, which has a characteristic absorption spectrum due to its distinct electronic structure from the neutral Cz molecule.[70–72] Hence, photoexcitation of the sub-gap state involves the transfer of an electron from Cz to the CB of the PbI$_4^{2-}$ layer, leaving a hole behind on Cz. The temporal compression associated with generating the broad-band pulse allows us to track the rise kinetics of the CT process with sub-15 fs temporal precision as demonstrated in Figure 2c. As the rise kinetics are convoluted with the coherent artifact (Figure S8) we conclude that the Cz$^{+\cdot}$ species is generated directly upon photoexcitation of the sub-gap state. Therefore, these collective observations revealing a sub-gap CT state provide the first unambiguous evidence of a direct contribution of the organic molecule to the optical properties of a perovskite. Moreover, the Cz-C$_3$ spacer molecule appears to be not only optically active, but also opto-electronically active as seen in the EQE spectrum, which is in stark contrast with the electronically inert behavior of the PEA spacer. Finally, we note that there is no Cz$^{+\cdot}$ species present in the ground state due to the absence of its near-IR absorption peak in the PDS spectrum (Figure 2a).

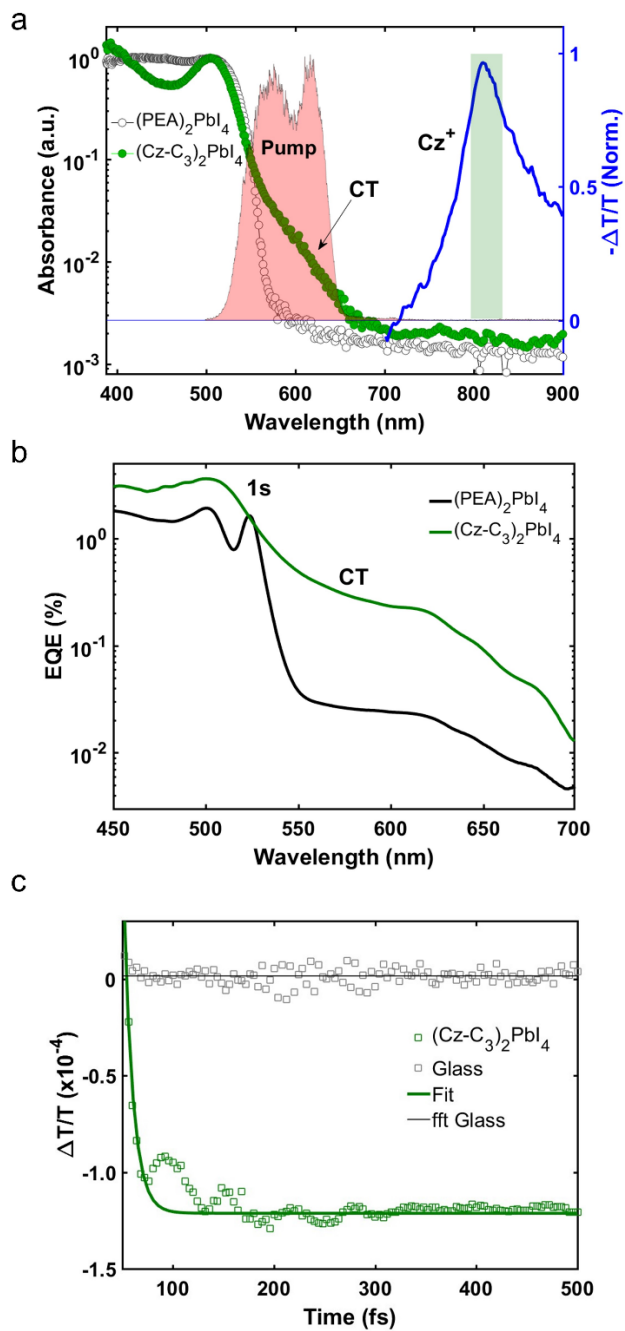

Figure 2. Optical organic-inorganic charge transfer in (Cz-C$_3$)$_2$PbI$_4$. (a) Photothermal deflection and transient absorption spectra. A sub-gap state is revealed in the photothermal deflection spectrum of (Cz-C$_3$)$_2$PbI$_4$ at ~600 nm, which is absent in (PEA)$_2$PbI$_4$. This state is excited with a broad-band pump pulse centered at 600 nm (red area, 25 μJ/cm$^2$) and the photoinduced absorption band of Cz$^{+\cdot}$ (blue spectrum) appears in the near-IR. This spectrum was integrated from 500-3000 fs. (b) External Quantum Efficiency spectrum of (Cz-C$_3$)$_2$PbI$_4$ and (PEA)$_2$PbI$_4$ photovoltaic devices. Note the different x-axis scale for b than a. (c) femtosecond rise kinetics of Cz$^{+\cdot}$ spectrally integrated from 790-820 nm (see green box in Figure 2a). The rise component in the kinetics is IRF (15 fs)-limited, as determined by the temporal pulse width of the pump (Figure S8) and is convoluted with the rise in glass (grey squares). A fast-Fourier transformation (fft) is shown for the latter as well, indicating that the corresponding signal is on the noise level.

The PDS spectra for (Cz-C$_4$)$_2$PbI$_4$ and (Cz-C$_5$)$_2$PbI$_4$ thin films are shown in Figure S7. Whereas (Cz-C$_4$)$_2$PbI$_4$ does not display any observable sub-gap state absorption, there is again a small shoulder around 600 nm in (Cz-C$_5$)$_2$PbI$_4$ suggesting a weak CT transition. Although less evident than in the C$_3$ analogue due to the low signal-to-noise ratio in the fs-TA spectrum of this C$_5$ material, the rise kinetics of the Cz$^{+\cdot}$ photoinduced absorption band is again convoluted with the coherent artifact, allowing us to assign the sub-gap absorption to a CT state in (Cz-C$_5$)$_2$PbI$_4$ (Figure S8).

From the slope of the sub-gap absorption tail, we derive Urbach energies,[73] indicating a higher degree of electronic disorder in (Cz-C$_4$)$_2$PbI$_4$ and (Cz-C$_5$)$_2$PbI$_4$ compared to (PEA)$_2$PbI$_4$ (Figure S7), which could be relevant for (defect-assisted) nonradiative decay. We do note that the Urbach energy (33 meV) in (Cz-C$_4$)$_2$PbI$_4$ and (Cz-C$_5$)$_2$PbI$_4$ is still close to room temperature thermal disorder (25 meV). As the sub-gap absorption tail is overlapping with the CT state in (Cz-C$_3$)$_2$PbI$_4$, we could not provide an Urbach energy for this material. Finally, we note the presence of a broad absorption band in the IR for (Cz-C$_5$)$_2$PbI$_4$, which could indicate the presence of mid-gap trap states or Cz-dimer formation.[74]

**Photoinduced Charge Transfer and Excited State Dynamics**

We now further study the excited state dynamics of (Cz-C$_i$)$_2$PbI$_4$ and (PEA)$_2$PbI$_4$ with TA spectroscopy to understand how the electronic structure of the organic ligand controls the recombination kinetics. We focus on (Cz-C$_4$)$_2$PbI$_4$ here as a representative system for (Cz-C$_i$)$_2$PbI$_4$. The TA spectra for a (Cz-C$_4$)$_2$PbI$_4$ thin film excited at 400 nm at several time delays (0.1-1000 ps) are shown in Figure 3a. The transient response in the spectral region of the excitonic transition (430-540 nm) is qualitatively similar to (PEA)$_2$PbI$_4$ (Figure S9), characterized by the spectral features labelled α, β and γ commonly observed in 2D perovskites.[75–77] Feature β is associated with a positive ΔT/T, i.e., an enhanced transmission at that particular wavelength. This "bleach" is the result of phase space filling due to photoexcitation. At the high-energy side of β is a negative ΔT/T feature, γ, which is the result of hot-carrier cooling. Lastly, on the low-energy side, another negative ΔT/T feature appears on the ps timescale, which is often associated with band-gap renormalization,[78] although this assignment remains controversial.[75] The ps-TA spectra in the blue spectral region for both (Cz-C$_3$)$_2$PbI$_4$ and (Cz-C$_5$)$_2$PbI$_4$ (Figure S10) again show the characteristic features for 2D perovskites as discussed above for (Cz-C$_4$)$_2$PbI$_4$ (Figure 3a). Feature α, however, in (Cz-C$_3$)$_2$PbI$_4$ has a positive ΔT/T, which is consistent with filling of the sub-gap CT states observed with PDS (Figure 2). Note that for (Cz-C$_5$)$_2$PbI$_4$ a much smaller initial carrier density $n_0$ (Section S1) is required to obtain the same initial ΔT/T at β indicating that on a sub-ps timescale a larger fraction of the exciton population has already decayed for (Cz-C$_3$)$_2$PbI$_4$ and (Cz-C$_4$)$_2$PbI$_4$ than for (Cz-C$_5$)$_2$PbI$_4$.

As demonstrated with fs-TA spectroscopy above for (Cz-C$_3$)$_2$PbI$_4$ (Figure 2a), the near-IR spectral region shows a broad photoinduced absorption band with a maximum at 800 nm characteristic for Cz$^{+\cdot}$, and this feature is completely absent in (PEA)$_2$PbI$_4$ (Figure 3b). However, whereas in the experiment described above the CT state was populated directly upon photoexcitation in the sub-gap region where organic orbitals contribute to the joint density of states, here we excite at 400 nm. The joint density of states is dominated by PbI$_4^{2-}$ orbitals rather than Cz orbitals at this wavelength, as it is energetically far above the CT state, yet below the first excited state of Cz-C$_4$ (Figure S3 and S4), hence generating PbI$_4^{2-}$ localized excitons. Therefore, in this case, it is hypothesized that Cz$^{+\cdot}$ is generated through photoinduced hole transfer from the photoexcited perovskite layer. The photoinduced hole transfer rate is on the order of the temporal resolution of the set-up (≤200 fs, see inset Figure 3b) and can therefore not be resolved with the current set-up. We note that photoinduced energy transfer from PbI$_4^{2-}$ to Cz may be excluded as a decay channel, as both the singlet and triplet states of the Cz molecules are energetically inaccessible.[79]

The appearance of photoinduced hole transfer in (Cz-C$_4$)$_2$PbI$_4$ is again consistent with the proposed band diagram (Figure 1b) where the HOMO-Cz lies energetically above the VBM of the PbI$_4^{2-}$ layer, hence providing an energetic driving force for hole transfer. It is also consistent with the detection of radicals on carbazole in a preliminary light-induced EPR study of (Cz-C$_5$)$_2$PbI$_4$ films conducted by some of the authors.[23] Moreover, it rationalizes the weak PL (Figure 1f) as the hole transfer process results in the electron and hole wavefunctions being spatially separated, giving rise to a reduced radiative decay probability. This observation is similar to the CT-induced PL quenching observed in napthalenediimide-based[21,22] and thiophene-based 2D perovskites,[20] and in a thiol-coupled 2D perovskite.[80] In contrast, the electron and hole remain strongly bound and localized on the inorganic layer in (PEA)$_2$PbI$_4$, resulting in efficient PL at the exciton transition.[81] Although it is tempting to relate the absorption of Cz$^{+\cdot}$ at 800 nm to the broad emission feature at 750 nm (Figure S5), it is unlikely that the latter corresponds to interlayer CT emission as the HOMO-Cz would have to lie energetically in the mid-gap region in order to result in emission at this wavelength. As mentioned earlier, this emission feature is instead likely related to phenomena that are more generally observed for 2D perovskites, such as deep traps or self-trapped excitons.

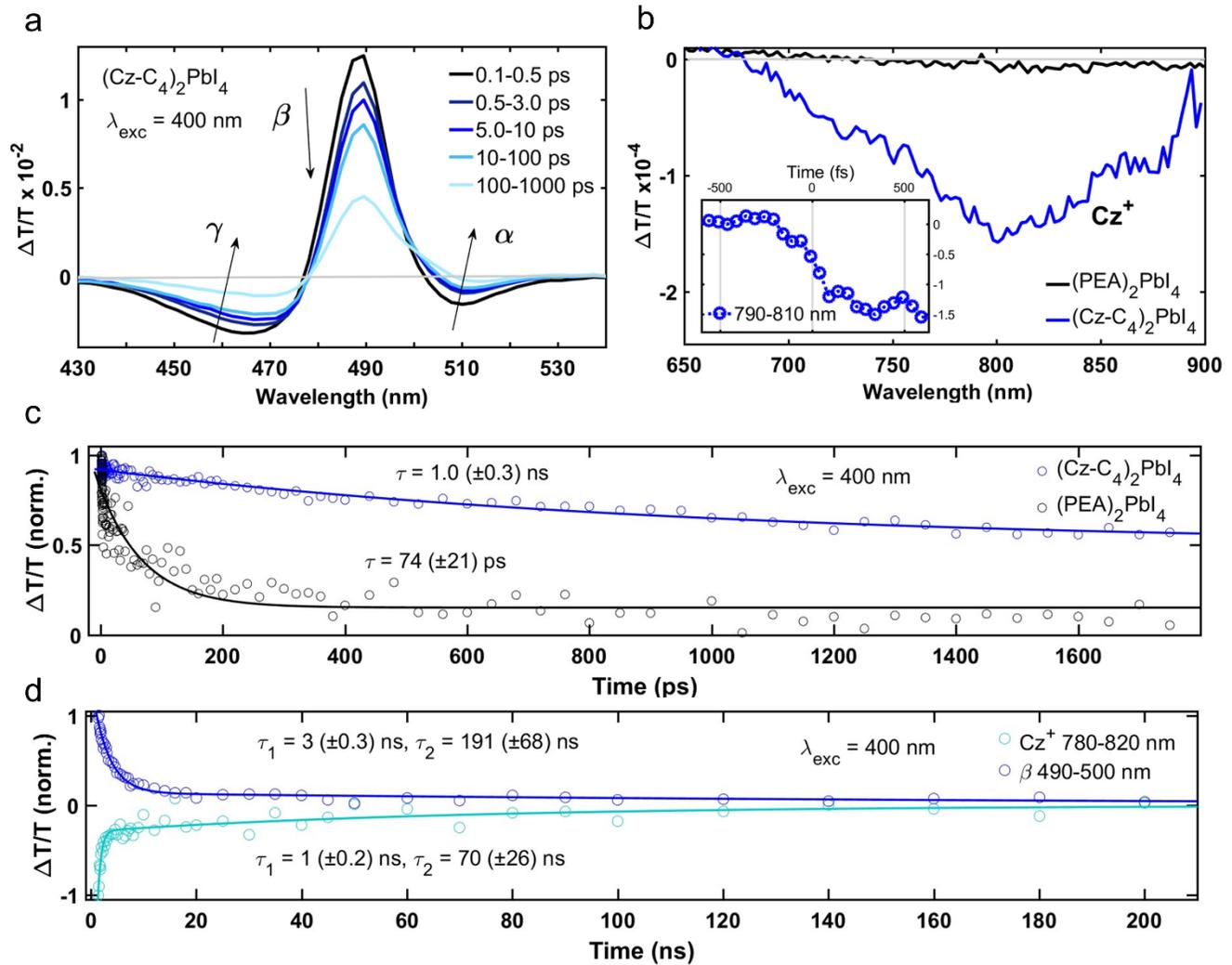

**Figure 3.** Excited state dynamics of a (Cz-C$_4$)$_2$PbI$_4$ thin film studied with ps- and ns-transient absorption spectroscopy. (a) transient absorption spectra integrated over different time regimes, excited at 400 nm (5.6 µJ/cm$^2$). The same general features as in the transient absorption spectra of (PEA)$_2$PbI$_4$ (Figure S9) are labelled with α, β and γ and are discussed in the text. (b) transient absorption spectra probed in the red integrated over 1-3 ps for (PEA)$_2$PbI$_4$ and (Cz-C$_4$)$_2$PbI$_4$ thin films. The inset shows the kinetics of the ultrafast rise of the photoinduced absorption band at 800 nm, assigned to the absorption of Cz$^{+\cdot}$. A higher fluence of 13.5 µJ/cm$^2$ was used to enhance the Cz$^{+\cdot}$ signal. (c) kinetics of the exciton bleach (β) for (PEA)$_2$PbI$_4$ and (Cz-C$_4$)$_2$PbI$_4$ thin films with the same carrier density 1.5x10$^{18}$ cm$^{-3}$ normalized at 1 ps. Mono-exponential fits are also indicated. (d) Normalized nanosecond kinetics of the exciton bleach and photoinduced absorption of a (Cz-C$_4$)$_2$PbI$_4$ thin film excited at 400 nm 3.2x10$^{19}$ cm$^{-3}$. Bi-exponential fits are also indicated.

We track the decay of β in both (Cz-C$_4$)$_2$PbI$_4$ and (PEA)$_2$PbI$_4$ on a ps timescale to understand the effect of photoinduced CT on the carrier dynamics (Figure 3c). Although it might be expected that CT would result in a faster decay of β due to the reduction of bleaching carriers, as the CT process is ultrafast (i.e. sub-ps), β does in fact not decay faster in (Cz-C$_4$)$_2$PbI$_4$ on a ps timescale. On the contrary, we observe a significantly longer carrier lifetime (Cz-C$_4$)$_2$PbI$_4$ (τ=1.0 ns) compared to (PEA)$_2$PbI$_4$ (τ=74 ps) as indicated by the mono-exponential fits.[82] To more accurately capture the excited state decay in (Cz-C$_4$)$_2$PbI$_4$, we also perform nanosecond TA spectroscopy (Figure 3d), revealing time constants (relative amplitudes) of τ$_1$ = 3 ns (90%) and τ$_2$ = 191 ns (10%) for the decay of β through bi-exponential fitting. This decay represents both the population of PbI$_4^{2-}$ localized excitons, as well as, depending on the quantum yield of the CT process (*vide infra*), a certain population of excitons formerly localized on the PbI$_4^{2-}$ layer that has been converted into a combination of Cz localized holes, which induces the absorption at 800 nm, and PbI$_4^{2-}$ localized electrons. The dominant factor in the >10-fold extension of the excited state lifetime in (Cz-C$_4$)$_2$PbI$_4$ compared to (PEA)$_2$PbI$_4$ is the reduction in PL efficiency due to the spatial separation of the electrons and holes, as manifested in the extremely weak PL (Figure 1f). Photoexcited carriers in (Cz-C$_4$)$_2$PbI$_4$ must therefore predominantly decay nonradiatively, either facilitated by strong exciton-phonon coupling or traps. It is unlikely that a more significant (trap-assisted) nonradiative decay pathway explains the faster decay for (PEA)$_2$PbI$_4$ compared to (Cz-C$_4$)$_2$PbI$_4$, as the (PEA)$_2$PbI$_4$ film was found to be of higher electronic quality based on the smaller Urbach energy (Figure S7). Furthermore, even though time-correlated single-photon counting (TCSPC) reveals a nanosecond decay component in the decay of the PL of (PEA)$_2$PbI$_4$, it is consistently much shorter lived than both the blue and red PL of (Cz-C$_4$)$_2$PbI$_4$, the latter even displaying a carrier lifetime component on the order of microseconds (Figure S5 and S15).

The decay of Cz$^{+\cdot}$ was also fitted to a bi-exponential function with time constants (relative amplitudes) of $\tau_1$ = 1 ns (95%) and $\tau_2$ = 70 ns (5%) (Figure 3d) and follows monomolecular kinetics as evident from its fluence-independent behavior (Figure S11), implying that bimolecular processes including exciton-exciton annihilation do not play a role for the fluences employed here. Therefore, we may conclude that the recombination process is geminate,[83] suggesting that the PbI$_4$$^{2-}$ localized electron and Cz localized hole form an excitonic CT state as they remain coulombically bound.[84] Furthermore, the decay behavior of β is also fluence-independent, as it is described by a combination of CT excitons and PbI$_4$$^{2-}$ localized excitons, both following monomolecular kinetics. This is further supported by the fluence-independent decay of both the blue and red emitting species (Figure S16).

The presence of long-lived carriers in (Cz-C$_3$)$_2$PbI$_4$ and (Cz-C$_5$)$_2$PbI$_4$ are also evident (Figure S11 and S15), akin to (Cz-C$_4$)$_2$PbI$_4$. The decays of β and Cz$^{+\cdot}$ are again fluence independent for (Cz-C$_3$)$_2$PbI$_4$. In contrast, their decay reveals a fluence dependence in (Cz-C$_5$)$_2$PbI$_4$, demonstrating bimolecular kinetics and therefore implying that the electron-hole recombination process is non-geminate.[85,86] The differences between non-geminate and geminate recombination pathways in (Cz-C$_5$)$_2$PbI$_4$ and (Cz-C$_{3,4}$)$_2$PbI$_4$, respectively, is potentially facilitated by the larger electron-hole separation imposed by the larger distance between the PbI$_4$$^{2-}$ layers and Cz ligands. Furthermore, the larger electron-hole distance in (Cz-C$_5$)$_2$PbI$_4$ results in the longest nanosecond lifetimes for every initial excitation density ($n_0$) investigated. A representative comparison for one $n_0$ is shown in Figure S12. However, it should be noted that the potential presence of mid-gap trap states as implied by the broad IR absorption band could significantly complicate the photophysics and result in bimolecular decay as well (Figure S7). From the above, it is clear that the interlayer CT induced by the electro-active spacer molecule results in significantly distinct excited state dynamics and kinetics.

**Tuning the Charge Transfer Quantum Yield**

We now aim to quantify the charge transfer (CT) quantum yield (QY) to understand what factors control the CT efficiency. To do so, we first investigate the fluence and excitation wavelength dependence of photoinduced CT. The CT QY, denoted as $\varphi_{CT}$, is defined as the ratio of Cz$^{+\cdot}$ molecules generated per unit volume (which is related to the magnitude of the TA signal) to the number of absorbed photons per unit volume, $n_0$ (Section S1). As we are interested in the magnitude of the Cz$^{+\cdot}$ signal, we again investigate the near-IR spectral region (Figure 3b) and vary both $n_0$ and the excitation wavelength, $\lambda_{exc}$, first by focusing on (Cz-C$_4$)$_2$PbI$_4$. We acquire a $\varphi_{CT}$ value of 11% for $\lambda_{exc}$ = 400 nm. This value does not depend on $n_0$, which is to be expected for a monomolecular CT process (Figure 4a). We note that in principle the $\varphi_{CT}$ could also be derived from the PL QY if it is assumed that each charge transfer event results in nonradiative decay and that there are no other nonradiative decay channels (e.g. trapping). Despite significant effort, an accurate PL QY could not be determined due to the low PL intensity of the samples (cf. Figure 1f).

Interestingly, resonant photoexcitation of the 1s state (495 nm) results in a two-fold increase in $\varphi_{CT}$ (Figure 4b). This result would not match with a photoinduced CT process involving an activation barrier, as in that case, a higher-energy photon should promote the transfer. Even if the CT would be barrierless, it is non-trivial to explain why a higher-energy photon reduces the efficiency. There might, however, be several possible explanations for this: i) photoexcitation

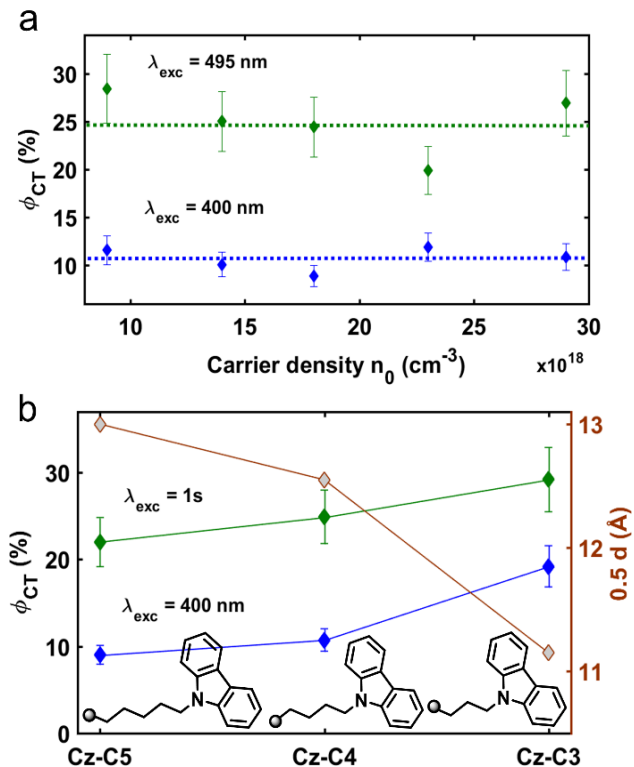

**Figure 4.** Tuning the charge transfer quantum yield, $\varphi_{CT}$. (a) Dependence of $\varphi_{CT}$ on excitation wavelength (green line for $\lambda_{exc}$ = 495 nm; blue line for $\lambda_{exc}$ = 400 nm) and carrier density. The dotted lines indicate the average over carrier density. (b) Dependence of $\varphi_{CT}$ on alkyl chain length and excitation wavelength (green line for resonant 1s excitation; blue line for $\lambda_{exc}$ = 400 nm). The 1s transition corresponds to 505 nm for (Cz-C$_3$)$_2$PbI$_4$ and (Cz-C$_5$)$_2$PbI$_4$ and 495 nm for (Cz-C$_4$)$_2$PbI$_4$ (Figure 1e). The values shown were averaged over multiple carrier densities (Figure S14). The red line with grey symbols indicates half the d-spacing which are derived from the XRD patterns (Figure 1d).

at 400 nm leads to additional loss pathways, either through more facile carrier trapping at defects[87] or a higher degree of carrier-carrier scattering[88], or ii) photoexcitation at 400 nm generates an exciton more localized on $PbI_4^{2-}$, whereas photoexcitation at 495 nm generates an exciton delocalized across the inorganic-organic interlayer, biasing the fate of the exciton towards an electron localized on $PbI_4^{2-}$ and a hole localized on Cz.

Next, we investigate the role of the distance between the inorganic layer and the Cz-core on $\varphi_{CT}$ by varying the alkyl chain length. Consistent with the $\lambda_{exc}$ dependent results for $(Cz-C_4)_2PbI_4$ (Figure 4a) the $\varphi_{CT}$ also increases for $(Cz-C_3)_2PbI_4$ and $(Cz-C_5)_2PbI_4$ when photoexcited at the 1s transition (green line) compared to photoexcitation at 400 nm (blue line). As for $i=4$, the $\varphi_{CT}$ does not depend on $n_0$ for $i=3$ and $i=5$ (Figure S14). However, for both 1s and 400 nm excitation, $\varphi_{CT}$ increases with a decreasing alkyl chain length (green and blue lines in Figure 4b). Hence, the efficiency of CT decreases monotonically with the distance between the inorganic layer and the Cz-core (red line in Figure 4b), which is estimated by taking half of the interplanar spacing determined from XRD (Figure 1d). This trend is also consistent with the decreasing PL intensity with decreasing alkyl chain length (Figure S5), when considering that the absorbance at 400 nm is similar for all films (Figure S3), although defects and self-trapped excitons could contribute to the trend in PL intensity. The largest $\varphi_{CT}$ for $(Cz-C_3)_2PbI_4$ is another manifestation of its strongest inorganic-organic interlayer coupling compared to its longer alkyl chain length analogues $i=4$ and $i=5$, as also evidenced through observation of the optical CT state (Figure 2).

The qualitative trend between interlayer CT efficiency and interlayer distance agrees with electron transfer theories, such as Marcus theory.[89] However, the exponential relation between distance and efficiency predicted by Marcus theory is not reproduced for the $(Cz-C_i)_2PbI_4$ series. This is most likely because not only the interlayer distance is changed from $i=3$ to $i=5$, but also the energetic position of relevant orbitals and therefore the electronic driving force for CT. Indeed, we observe spectral shifts in the 1s excitonic transition (Figure 1e), as well as the absorption spectra of Cz (Figure S3) and $Cz^{+\cdot}$ (as seen in the ps-TA spectra in Figure S13), clearly demonstrating that relevant orbitals for those transitions shift in energy, potentially having contributions from distortion of the $PbI_4^{2-}$ layer and the inductive effect imposed by the alkyl substituent.

### Vertical Charge Transport

To verify an improved interlayer electronic coupling on a longer length scale relevant to charge transport, we measured the electrical vertical charge transport in $(Cz-C_i)_2PbI_4$ and $(PEA)_2PbI_4$ devices (Figure 5 inset). The surface morphologies of the perovskite layers are smooth as visualized with AFM images (Figure S17). We note that due to the preferential growth of the 2D $PbI_4^{2-}$ layers parallel to the substrate (Figure 1d), we mainly track charge transport along the out-of-plane (OOP) direction in these vertical architecture devices. J-V curves of $(Cz-C_i)_2PbI_4$ and $(PEA)_2PbI_4$ devices are provided in Figure S18, with all devices exhibiting ohmic regimes ($J \propto V$) at low voltage. OOP conductivities ($\sigma_{OOP}$), derived from the intersect of the linear fit in this regime, are shown in Figure 5 (black circles). The $\sigma_{OOP}$ increases from $(PEA)_2PbI_4 < (Cz-C_5)_2PbI_4 < (Cz-C_4)_2PbI_4 < (Cz-C_3)_2PbI_4$, indicating that both the electro-active ligand as well as the decreasing alkyl chain length improve OOP charge transport. The enhanced $\sigma_{OOP}$ for $(Cz-C_3)_2PbI_4$ compared to $(PEA)_2PbI_4$ is consistent with its larger EQE values (Figure 2b). Interestingly, the $\sigma_{OOP}$ of $(Cz-C_3)_2PbI_4$ exceeds both the out-of-plane and in-plane conductivity ($1\cdot10^{-13}$ S/cm and $2\cdot10^{-11}$ S/cm, respectively) measured in a $(PEA)_2PbI_4$ single crystal.[90]

To eliminate contributions of different carrier concentrations to the OOP transport, we also determine out-of-plane charge carrier mobilities ($\mu_{OOP}$) from Child's law regime ($J \propto V^2$) using the appropriate device stack (Figure 5 inset) for space charge limited current (SCLC) (Section S2) measurements.[91] $\mu_{OOP,electron}$ increases in the order $(PEA)_2PbI_4 < (Cz-C_5)_2PbI_4 < (Cz-C_4)_2PbI_4 < (Cz-C_3)_2PbI_4$ (blue squares Figure 5), consistent with the trend in $\sigma_{OOP}$. Given the larger interplanar distance in the $(Cz-C_i)_2PbI_4$ films (Figure 1d) compared to $(PEA)_2PbI_4$, the larger $\mu_{OOP,electron}$ values in the former are induced by the electro-active Cz ligand, whereas the trend in increasing $\mu_{OOP,electron}$ from $(Cz-C_5)_2PbI_4$ to $(Cz-C_3)_2PbI_4$ is explained by the decreasing alkyl chain length and its corresponding shorter interplanar and $Cz-PbI_4^{2-}$ distances. The largest $\mu_{OOP,electron}$ value for $(Cz-C_3)_2PbI_4$ indicates the strongest *long-range* out-of-plane coupling and is

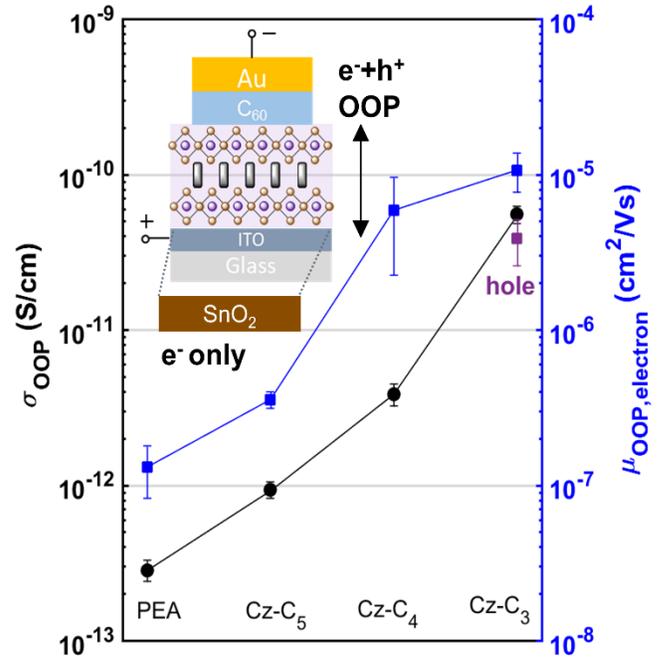

**Figure 5.** Vertical charge transport properties of $(Cz-C_i)_2PbI_4$ and $(PEA)_2PbI_4$ devices. Out-of-plane conductivities ($\sigma_{OOP}$, black circles) were measured on the device architecture shown in the inset. Out-of-plane electron mobilities ($\mu_{OOP,electron}$, blue squares) were measured on the same device having one extra $SnO_2$ layer on top of the ITO substrate. The $\mu_{OOP,hole}$ value for $(Cz-C_3)_2PbI_4$ is indicated with a purple square and was measured on a glass/ITO/2PACz/(Cz-$C_3)_2PbI_4$/PTAA/Au device. These values were extracted from J-V curves provided in Figure S18.

consistent with its largest *single* interlayer out-of-plane coupling, as evident from both the sub-gap CT state (Figure 2) as well as the largest CT QY observed with TA spectroscopy (Figure 4b).

The apparent synergy between long-range and short-range out-of-plane coupling may be explained by the reduced electron-hole Coulombic interaction induced by the single interlayer CT process. Furthermore, the same electronic reasons that drive interlayer CT, such as strong out-of-plane orbital overlap, are also important for out-of-plane charge transport. Considering a quantum well model,[92] the larger dielectric constant associated with the Cz-C$_i$ spacer molecules compared to PEA (Figure S19) decreases the activation barrier for tunneling, promoting both interlayer charge transfer and transport. Then, the increasing $\mu_{OOP,electron}$ from (Cz-C$_5$)$_2$PbI$_4$ < (Cz-C$_4$)$_2$PbI$_4$ < (Cz-C$_3$)$_2$PbI$_4$ is the result of a decreasing interlayer tunneling barrier width.

Due to the rapid current shortage of most of the prepared hole-selective devices, we were only able to determine $\mu_{OOP,hole}$ for (Cz-C$_3$)$_2$PbI$_4$. It is evident by comparing $\mu_{OOP,electron}$ and $\mu_{OOP,hole}$ for (Cz-C$_3$)$_2$PbI$_4$ that the long-range out-of-plane coupling has significant contributions from both electron and hole carriers, which is consistent with small out-of-plane electron and hole carrier masses predicted for 2D perovskites with electroactive spacer molecules.[35] The observation that hole carriers are moderately mobile in the out-of-plane direction despite the predicted band type II alignment indicates significant orbital hybridization between the organic and inorganic valence bands in this material.[10,93,94]

To the best of our knowledge, these are the first reported SCLC measurements on polycrystalline n=1 2D perovskite-based vertical devices. Furthermore, whereas out-of-plane[36] and in-plane (photo)conductivities[22] have been measured for 2D perovskites with electroactive ligands, out-of-plane mobilities were not determined yet. However, during the preparation of this manuscript, Sivula and coworkers published results of SCLC measurements on a <n>=5 Dion-Jacobson perovskite containing a naphthalenediimide derivative. The scarcity of SCLC-measured $\mu_{OOP}$ values for n=1 2D perovskites is most likely due to the extremely small value, often leading to shorting prior to reaching the SCLC regime.[95,96]

The electrical transport measurements presented herein clearly demonstrate out-of-plane electronic coupling between the organic and inorganic layers. However, it remains unclear in what charge transport regime the out-of-plane transport takes place, which depends on exciton (interlayer) delocalization.[97] Additionally, organic-inorganic hybrid orbitals could be relevant for strong out-of-plane coupling as well.[36] The mechanistic details associated with the enhanced vertical charge transport in (Cz-C$_i$)$_2$PbI$_4$ will be a subject of further study.

CONCLUSIONS

This study has revealed significant differences in optical properties and charge carrier dynamics for 2D perovskites incorporating either electronically active (Cz-C$_i$) or inactive (PEA) spacer molecules. Firstly, the direct observation of a sub-gap interlayer CT state in (Cz-C$_3$)$_2$PbI$_4$ reveals the optically active nature of the Cz molecule and a strong interlayer coupling. The distinct excited state dynamics of (Cz-C$_i$)$_2$PbI$_4$ is driven by ultrafast photoinduced hole transfer from the inorganic PbI$_4^{2-}$ layer to the Cz molecule, whereas the excited state dynamics in (PEA)$_2$PbI$_4$ is described by excitons confined to the PbI$_4^{2-}$ layers. The hole transfer process increases the electron-hole separation, which extends the carrier lifetime compared to (PEA)$_2$PbI$_4$ due to the decreased probability of radiative recombination. The efficiency of photoinduced hole transfer ($\varphi_{CT}$) was found to decrease monotonically with the distance between PbI$_4^{2-}$ layers and the Cz core, demonstrating design rules for tuning optoelectronic properties in 2D perovskites through organic ligand modification. We have validated such an approach by determining the impact of the electroactive organic cations on out-of-plane (vertical) charge carrier mobility. We observe i) an increased out-of-plane mobility in (Cz-C$_i$)$_2$PbI$_4$ compared to (PEA)$_2$PbI$_4$ due to the electroactive ligand and ii) an increased out-of-plane mobility in (Cz-C$_i$)$_2$PbI$_4$ with decreasing alkyl chain length (from *i*=5 to *i*=3). The improved out-of-plane mobility is another manifestation of the enhanced electronic coupling between the inorganic and organic layers and stimulates the potential usage of electroactive spacer molecules in perovskite photovoltaic cells as well as other optoelectronic devices with new functionality.

EXPERIMENTAL SECTION

**Materials.** All commercial chemicals and solvents were used without additional purification steps unless stated otherwise. PbI$_2$ (lead(II) iodide, 99.99%) was obtained from TCI. PEA (phenethylamine, 99%), 9H-carbazole (96%), 3-bromopropylammonium bromide (98%), di-t-butyl dicarbonate (97%), triethylamine (99%), HI (57% in water, distilled, unstabilized), and tri-n-butyl phosphate (98%) were purchased from Acros Organics. Sodium t-butoxide (97%) was purchased from Sigma-Aldrich. The dry THF (tetrahydrofuran) used during the synthesis of Cz-C$_3$I, and the dry DMF (dimethylformamide) used to prepare the perovskite precursor solutions were obtained from an in-house solvent purification system (MBRAUN SPS-800). All other solvents were purchased from Fisher Scientific.

**Deposition of 2D perovskite films.** The ammonium salt of the large organic cation was dissolved in DMF together with lead iodide in a 2:1 molar ratio (the concentration used depends on the experiment and are indicated below). The precursor solutions were stirred at 50 °C for 1 hour. All precursor solutions were subsequently filtered through a PTFE syringe filter (0.45 μm mesh). Substrates were cleaned through successive sonication steps in the following order of solvents (detergent water, deionized water, acetone, and isopropanol; 15 min for each step), followed by a UV-ozone treatment of 15 min. The precursor solutions were deposited as thin films by spin coating via a one-step method and annealed on a hot plate in a glovebox under a nitrogen atmosphere (<0.1 ppm of O$_2$, <0.1 ppm of H$_2$O). The spin parameters and the annealing temperature that were used depend on the technique for which the samples were prepared and on the perovskite composition and are indicated below.

The samples were kept in a glovebox and removed only for analysis.

For steady-state absorption and emission spectroscopy, XRD, and PDS, films were deposited using a precursor solution concentration of 0.5 M for $PbI_2$ and by spinning at 2000 rpm, 2000 rpm/s for 20s. For transient absorption spectroscopy, films were deposited using a precursor solution concentration of 0.04 M for $PbI_2$ and by spinning at 6000 rpm, 4000 rpm/s for 20s, to obtain the thinner films necessary for this experiment. The $(PEA)_2PbI_4$ films were annealed at 110 °C for 10 min, $(Cz-C_4)_2PbI_4$ and $(Cz-C_5)_2PbI_4$ films at 110 °C for 15 min, and $(Cz-C_3)_2PbI_4$ at 130 °C for 15 min.

**Steady-state Characterization.** *X-ray Diffraction.* X-ray diffraction patterns of thin films on quartz substrates were recorded on a Bruker D8 Advance Powder X-ray Diffractometer with $CuK_\alpha$ radiation at ambient temperature.

*Atomic Force Microscopy.* Surface topography and thickness of perovskite films are imaged with the Bruker Dimension Icon Pro atomic force microscope with a silicon tip on nitride lever (Bruker Scanasyst-Air cantilever with spring constant 0.4 N/m). Film thicknesses are assessed by scanning with peak force tapping mode across the depth of a scratch made on the film with a razor blade. All data are analysed with the WSxM 5.0 software.[98]

*Ultraviolet-Visible Absorption Spectroscopy.* Ultraviolet–visible absorption (UV–Vis) spectra were recorded on a Shimadzu UV–VIS–NIR Spectrophotometer UV-3600Plus. A glass substrate was used as a blank.

*Photothermal Deflection Spectroscopy.* For photothermal deflection spectroscopy (PDS) experiments the thin films (~500 nm) were spin-coated on Spectrosil® 2000 fused silica substrates. The samples were excited with a monochromatic pump beam coming from a tuneable light source consisting of a quartz tungsten halogen lamp and a grating monochromator, mechanically modulated at 10 Hz. A portion of the absorbed light energy converts into heat through nonradiative recombination, producing an alternating temperature gradient at the sample surface. With the samples immersed in a liquid with a high thermo-optic coefficient (3M™ Fluorinert™ FC-72), which creates a thermal lensing effect around the excitation spot, they were probed with a continuous wave probe laser beam (670 nm) passed parallel to the perovskite layer surface. The beam deflection was detected with a quadrant photodiode and demodulated with a lock-in amplifier. PDS enables the measurement of a signal proportional to absorbance with a high dynamic range, while remaining insensitive to light scattering and other unwanted effects present in UV–VIS spectroscopy.

*Steady-State Photoluminescence Spectroscopy.* Photoluminescence spectra of $(Cz-C_i)I$ thin films on quartz substrates were recorded on a FLS1000 with a monochromatic Xe lamp as the excitation source and a photomultiplier tube as a detector. Because of the extremely weak photoluminescence of the $(Cz-C_i)_2PbI_4$ films, the photoluminescence spectra of 2D perovskite films were recorded using an intensified charge-coupled detector (iCCD). The experimental details are provided below (time-resolved photoluminescence spectroscopy).

**Time-resolved Spectroscopy.** *Transient Absorption spectroscopy.* Transient absorption (TA) spectroscopy is a technique that measures the change in transmission after photoexcitation with a pump beam. When the pump-probe delay time is systematically varied, the kinetics of the excited-state can be determined. Whereas for all the ps- and fs-TA measurements the pump-probe time delay was mechanically modulated, the nanosecond TA measurements made use of an electronic delay generator. In both cases, the pump pulse is blocked by a chopper wheel which rotates at half the frequency of the laser repetition rate to record transmission spectra with the pump on and off repeatedly.

Picosecond (ps) TA measurements were performed on two setups with different probe spectra to span a broad wavelength range. Firstly, a broad ultraviolet-visible spectrum ranging from 400–600 nm was generated by pumping a $CaF_2$ crystal with the output of a 800 nm Ti:Sapphire laser (Spectra Physics Solstice Ace, 7 W, 1 kHz repetition rate, 100 fs) and collimating after it. The same output was used to generate a pump wavelength of 400 nm by second harmonic generation (SHG) through a β-barium borate crystal. On the second setup, a chirped white light continuum (WLC) spectrum ranging from 530–950 nm was generated by pumping a YAG crystal with the output of a 1030 nm Yb:KGW amplifier laser (Light Conversion Pharos, 14.5 W, 38 kHz repetition rate, 200 fs). Various pump wavelengths were generated through a TOPAS optical parametric amplifier with the 1030 nm seed pulse.

The fs-TA experiments were performed in a home-built setup with sub-15 fs temporal resolution using the same broad visible spectrum as a probe pulse. The sub-15 fs pump pulse was generated via non-collinear optical parametric amplification (NOPA) as reported previously.[99] An automatic harmonic generator (Light Conversion HIRO) was used to generate the third harmonic (343 nm) pulse required to pump the NOPAs. For the band selective experiment, a NOPA seeded by 1030-WLC and amplified by the third harmonic (343 nm) was used to generate a sub-15 fs pulse (as shown in Figure S8). The pump pulses were compressed using a combination of chirped mirrors and wedge prisms (Layertec). The spatio-temporal profile of the pulses was measured through second-harmonic generation frequency-resolved optical gating (SHG-FROG). To generate differential transmission spectra, a chopper wheel modulated the pump beam at 9 kHz. The pump-probe delay was set by a computer-controlled piezoelectric translation stage (PhysikInstrumente) with a step size of 4 fs, and the pump and probe polarizations were set to be parallel. The transmitted probe was recorded with a grating spectrometer equipped with a Silicon camera (Entwicklungsbüro Stresing) operating at 38 kHz and a 550 nm blazed grating. Thin film samples were prepared on 170±5 μm quartz substrate, and pulse compression was performed by placing the same substrate in the beam path of FROG to compensate for the dispersion effect produced by the cover slip.

*Time-resolved Photoluminescence Spectroscopy.* We employed an electronically gated iCCD camera (Andor iStar DH740 CCI-010) coupled with a calibrated grating spectrometer (Andor SR303i) to capture transient photoluminescence spectra at nanosecond timescales. The same pump pulse used for picosecond TA measurements in the blue

spectral region was utilized. To prevent scattered laser signals from interfering with the spectrometer, we used a 450 nm long-pass filter (Edmund Optics). We obtained the temporal evolution of the PL emission by varying the iCCD delay with respect to the excitation pulse, with a minimum gate width of 5 ns.

*Time-correlated Single-photon counting.* A pulsed laser (PicoQuant LDH-P-C-400B) operating at 100 kHz was used to excite thin films at 407 nm with a pulse energy of 8.5 pJ. The emitted photons were filtered using different combinations of long-pass and short-pass filters (495LP + 575SP, 650LP + 850SP) to resolve the two spectrally distinct emission features. A SPAD (Micro Photon Devices PDM) was used in conjunction with timing electronics from Picoquant (TimeHarp 260) to complete the TCSPC system. The instrument response function was determined by collecting scattered light from scratched glass, with a resultant time resolution of around 700 ps.

*Device preparation.* 0.15M $(A)_2PbI_4$ precursor solutions were prepared by dissolving AI and $PbI_2$ powders in a co-solvent of DMF-DMSO (4:1 volume ratio) and stirred at room temperature for 30 minutes and filtered by 0.2 μm pore-size PTFE filter. Then 20 μL of the solution was spin-coated on an ITO substrate at 5000 rpm/s for 30s and then annealed at 100 °C for 10 mins. The above processes were carried out in nitrogen-filled gloveboxes. Finally, 20 nm of $C_{60}$ was thermally evaporated onto the perovskite film followed by 40 nm of Au using a shadow mask. For the photovoltaic devices used to measure EQE, as well as the hole-selective vertical transport devices, a 1 mM solution of 2PACz (TCI) in anhydrous ethanol was spin-coated on top of the ITO substrate at 3000 rpm (5s ramp) for 30s, followed by annealing for 10 minutes at 100°C. For the hole-selective devices (Figure 5), instead of $C_{60}$ evaporation, PTAA (EM Index) solution (10 mg mL$^{-1}$ in toluene) doped with bis(trifluoromethylsulfonyl)imide lithium salt (Li-TFSI, 1.6 μL mL$^{-1}$ of a 1.8M solution in acetonitrile (ACN)) and 4-tert-butylpyridine (TBP; 2 μL mL$^{-1}$) was spun at 4000 rpm for 20s on top of the perovskite layer. For the electron-selective vertical transport devices (Figure 5), a 25 nm $SnO_2$ layer was deposited by atomic layer deposition (Picosun) on the ITO substrate. Tetrakis(dimethylamino)tin(IV) (TDMASn, EpiValence) was used as precursor and $H_2O$ as reactant. The precursor bubbler was heated to 75°C and the chamber to 120°C, the reactant vessel was kept at room temperature. The pulsing sequence consisted of a 0.8s pulse of TDMASn, 20s purge, 0.2s pulse of $H_2O$, 20s purge, resulting in a growth rate of 0.1 nm/cycle.

*External Quantum Efficiency.* EQE was measured using a Bentham PVE300 system in AC mode. A dual xenon short-arc lamp and a quartz halogen lamp were utilized as the light source, with a swingaway mirror set to 700 nm. A 10 × 10 mm Si reference cell was used to calibrate the power of the probe beam.

*Space-charge Limited Current.* Dark I-V and C-F measurements were carried out on a Desert TTP4 probe station by an Agilent 4155C Semiconductor Parameter Analyzer and Hewlett Packard 4192A LF Impedance Analyzer. Samples were loaded into the probe station chamber and pumped to high vacuum (<10$^{-5}$ mbar). I-V characteristics were measured in a pulsed mode with a scan speed of 100 mV/s.

Dark and light I-V characteristics for the photovoltaic device were collected using an Autolab PGSTAT302N (Metrohm) and an LED solar simulator (G2V Sunbrick Base-UV). An active area of 0.06 cm$^2$ was used. Devices were scanned at a scan speed of 100 mV/s.

## ASSOCIATED CONTENT

**Supporting Information**.
The Supporting Information is available free of charge at http://pubs.acs.org.
Additional experimental details, synthetic procedures, $^1$H-NMR and $^{13}$C-NMR spectra, AFM images, XRD patterns, UV–Vis spectra, PL spectra, PDS spectra, TA spectra and kinetics, TCSPC curves, TR-PL spectra, J-V and C-F curves, device parameters.

## AUTHOR INFORMATION


### Corresponding Author

**\*Akshay Rao** - Cavendish Laboratory, University of Cambridge, JJ Thomson Ave, Cambridge CB3 0HE, United Kingdom; https://orcid.org/0000-0003-4261-0766; Email: ar525@cam.ac.uk

**\*Samuel D. Stranks** - Department of Chemical Engineering & Biotechnology, University of Cambridge, Philippa Fawcett Drive, Cambridge CB3 0AS, U.K.; Cavendish Laboratory, University of Cambridge, JJ Thomson Avenue, Cambridge CB3 0HE, U.K.; https://orcid.org/0000-0002-8303-7292; Email: sds65@cam.ac.uk

### Author Contributions

†Y. B., W.T.M.V.G. and Y.Z. contributed equally to this work.



### Funding Sources

Y.B. acknowledges the Winton Programme for Physics of Sustainability for funding. Y.Z. acknowledges financial support from Cambridge University Postgraduate Hardship Funding, Cambridge University PGR Covid-19 Assistance Scheme and EPSRC Centre for Doctoral Training in Graphene Technology. S.D.S. acknowledges the Royal Society and Tata Group (grant no. UF150033). The work has received funding from the European Research Council under the European Union's Horizon 2020 research and innovation program (HYPERION, grant agreement no. 756962; PEROVSCI, 957513). The authors acknowledge funding from the EPSRC (EP/S030638/1, EP/T02030X/1, EP/V027131/1). The special research fund (BOF) of UHasselt is acknowledged for the funding of the research activities of W.T.M.V.G. with the funding of a temporary postdoctoral fellowship (BOF22PD01). A.M. and I.D. acknowledge the FWO for the funding of their fundamental research Ph.D. grant (1115721N) and strategic basic research Ph.D. grant (1S31323N), respectively. L.L. and D.V. acknowledge the FWO for the funding of the SBO project PROCEED (FWOS002019N), and the senior FWO research projects G043320N and G0A8723N. S.L. is funded by the PROCEED project. S. J. Z. acknowledges support from the Polish National Agency for Academic Exchange within the Bekker program (grant no. PPN/BEK/2020/1/00264/U/00001). K.D acknowledges the financial support of Cambridge Trust in the form of Cambridge India Ramanujan Scholarship and Cambridge Philosophical Society for a research studentship.




Notes
The authors declare no competing financial interest.

# Tailoring Interlayer Charge Transfer Dynamics in 2D Perovskites with Electroactive Spacer Molecules


Yorrick Boeije,[1,2,†] Wouter T.M. Van Gompel,[3,†] Youcheng Zhang,[2,4,†] Pratyush Ghosh,[2] Szymon J. Zelewski,[1,2] Arthur Maufort,[3] Bart Roose,[1] Zher Ying Ooi,[1] Rituparno Chowdhury,[2] Ilan Devroey,[3] Stijn Lenaers,[3] Alasdair Tew,[2] Linjie Dai,[1,2] Krishanu Dey,[2] Hayden Salway,[1] Richard H. Friend,[2] Henning Sirringhaus,[2] Laurence Lutsen,[3] Dirk Vanderzande,[3] Akshay Rao,[2,*] Samuel D. Stranks[1,2,*]

[1]Department of Chemical Engineering and Biotechnology, University of Cambridge, Philippa Fawcett Drive, Cambridge, CB3 0AS, UK.

[2]Department of Physics, Cavendish Laboratory, University of Cambridge, JJ Thomson Ave, Cambridge, CB3 0HE, UK.

[3]Hasselt University, Institute for Materials Research (IMO-IMOMEC), Hybrid Materials Design (HyMaD), Martelarenlaan 42, B-3500 Hasselt, Belgium.

[4]Cambridge Graphene Centre, Department of Engineering, University of Cambridge, JJ Thomson Ave, Cambridge, CB3 0FA, UK.

Email: ar525@cam.ac.uk (A.R.), sds65@cam.ac.uk (S. D. S.)






**Synthesis of organic ammonium salts**

PEAI (phenethylammonium iodide) salt was synthesized as follows: phenethyl amine (1.04 mL, 8.25 mmol) was dissolved in ethanol (30 mL), and the flask was subsequently wrapped in aluminium foil. HI (57 wt% in water, unstabilized) was extracted three times with a chloroform/tri-*n*-butyl phosphate mixture (10:1, V/V).[1] Afterwards, the purified HI (1.14 mL, 8.67 mmol) was added to the reaction flask. The resulting mixture was left to stir at room temperature for 2 hours. After 2 h, a portion of ethanol was evaporated under reduced pressure. The residue was then precipitated in a large amount of cooled diethyl ether. The precipitate was collected by filtration and washed several times with diethyl ether. The product was obtained as a white solid (1.65 g, 6,62 mmol; 80% yield). $^1$H NMR (400 MHz, DMSO-$d_6$) δ 7.74 (s (br), 3H), 7.37-7.32 (m, 2H), 7.28-7.25 (m, 3H), 3.07-3.02 (m, 2H), 2.88-2.82 (m, 2H).

Cz-C$_3$I (3-(9*H*-carbazol-9-yl)propylammonium iodide) salt was synthesized as follows:

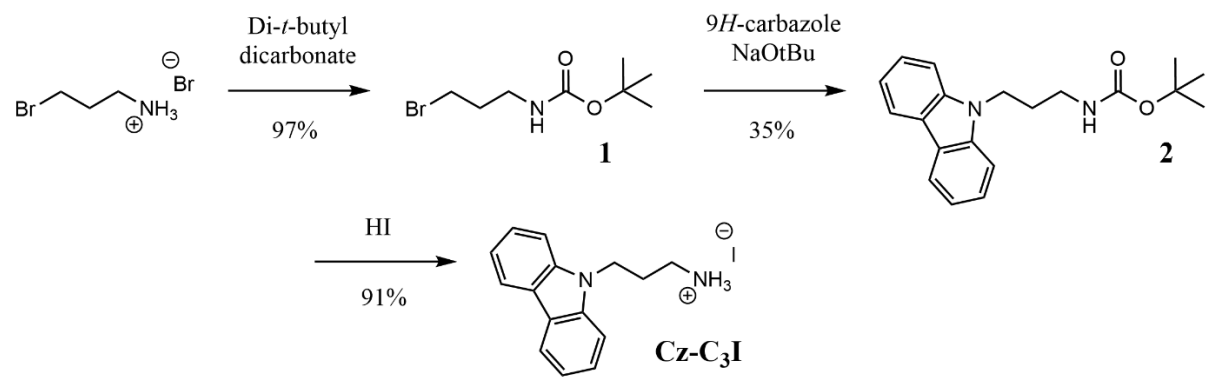

**3-bromopropylamine, Boc protected (1)**

3-bromopropylammonium bromide (5.00 g, 22.8 mmol) was dissolved in 100 mL dichloromethane and 3.50 mL triethylamine. Di-*t*-butyl dicarbonate (4.23 g, 19.4 mmol) was then added in one portion, after which carbon dioxide gas bubbles formed. The flask was sealed with a septum, which was perforated with two needles to allow carbon dioxide to escape. After 24 h at ambient temperature, the reaction mixture was extracted with water. The organic fraction was then dried, filtered, and concentrated to yield the target compound as a colourless oil (4.49 g, 97% yield). $^1$H NMR (400 MHz, Chloroform-*d*) δ 4.66 (s, 1H), 3.44 (t, J = 6.5 Hz, 2H), 3.27 (m, J = 6.5 Hz, 2H), 2.05 (p, J = 6.5 Hz, 2H), 1.44 (s, 9H). $^{13}$C NMR (101 MHz, Chloroform-*d*) δ 156.07, 79.50, 39.05, 32.76, 30.93, 28.47. GC-MS: m/z = 237 & 239 (target) and 181 & 183 (Boc deprotected target).



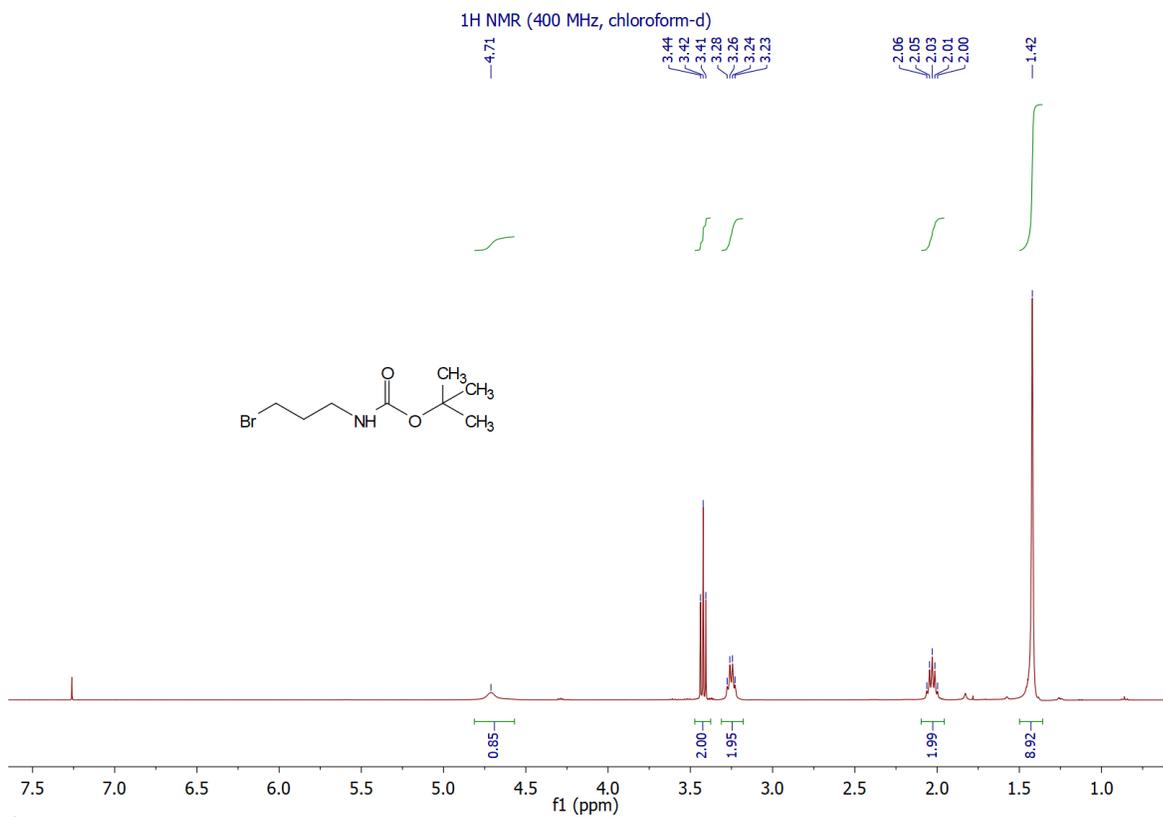

¹H NMR spectrum of 3-bromopropylamine, Boc protected (1) (residual internal CHCl₃ at 7.26 ppm).

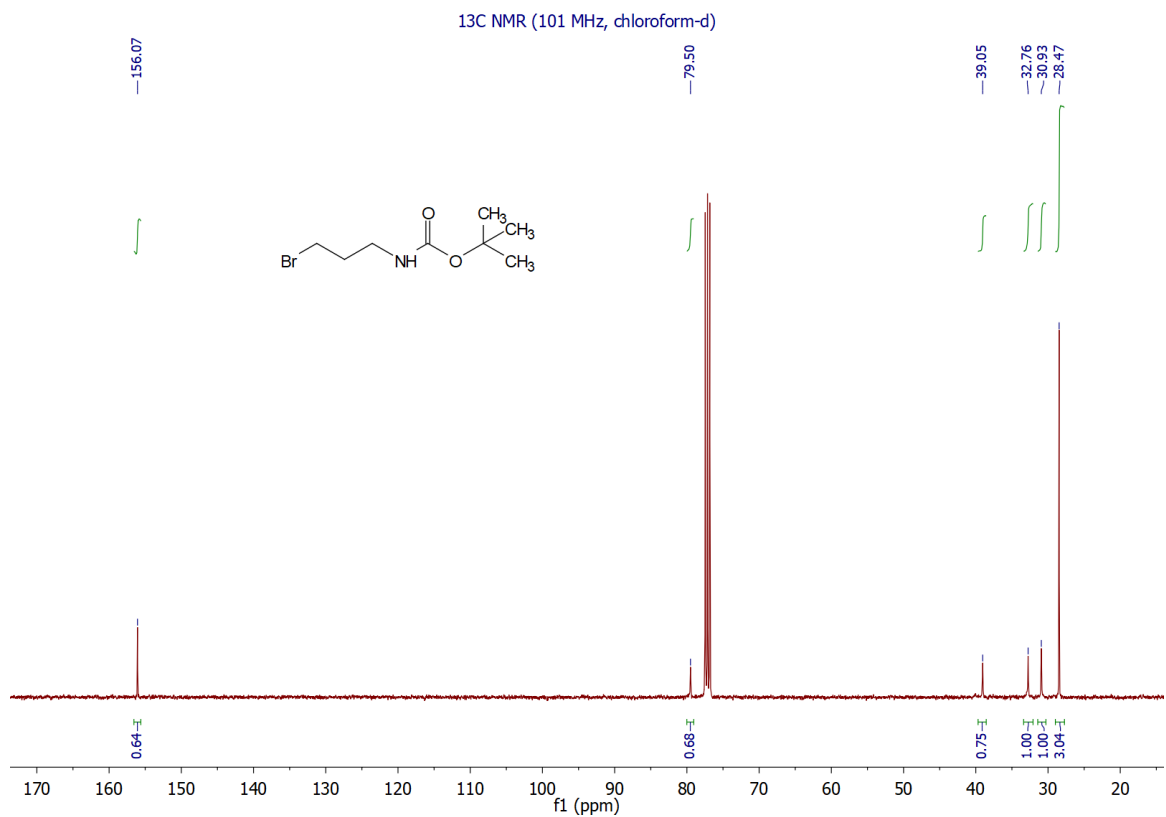



$^{13}$C NMR spectrum of 3-bromopropylamine, Boc protected (1) (solvent signal: CDCl$_3$ at 77 ppm).

### 3-(9H-carbazol-9-yl)propylamine, Boc protected (2)

9H-carbazole (1.99 g, 12.0 mmol) and sodium *t*-butoxide (1.17 g, 12.0 mmol) were dissolved in 25 mL dry THF under Ar atmosphere. **1** (7.15 g, 29.9 mmol) was added, and the resulting mixture was left to react at 55°C under Ar for 6 h. The mixture was then quenched with 1 M NH$_4$Cl and extracted with chloroform. The chloroform fractions were washed with water, dried, filtered, and concentrated. The resulting crude compound was purified through column chromatography (gradient elution from chloroform to chloroform/ethyl acetate 9:1) and vacuum distillation (Kugelrohr) to yield the target compound as a white powder (1.37 g, 35% yield). $^1$H NMR (400 MHz, Chloroform-*d*) δ 8.10 (dt, J = 7.7, 1.0 Hz, 2H), 7.47 (ddd, J = 8.2, 7.0, 1.2 Hz, 2H), 7.40 (dt, J = 8.2, 0.9 Hz, 2H), 7.23 (ddd, J = 8.0, 7.0, 1.1 Hz, 2H), 4.51 (s, 1H), 4.37 (t, J = 7.0 Hz, 2H), 3.15 (m, J = 7.3 Hz, 2H), 2.09 (p, J = 7.0 Hz, 2H), 1.43 (s, 9H). $^{13}$C NMR (101 MHz, Chloroform-*d*) δ 156.09, 140.31, 125.88, 123.01, 120.56, 119.09, 108.58, 79.53, 40.54, 38.62, 29.34, 28.48. GC-MS: m/z = 224 (Boc deprotected target).

$^1$H NMR spectrum of 3-(9H-carbazol-9-yl)propylamine, Boc protected (2) (residual internal CHCl$_3$ at 7.26 ppm; water at 1.56 ppm; ethyl acetate at 4.12 ppm, 2.03 ppm, and 1.25 ppm).

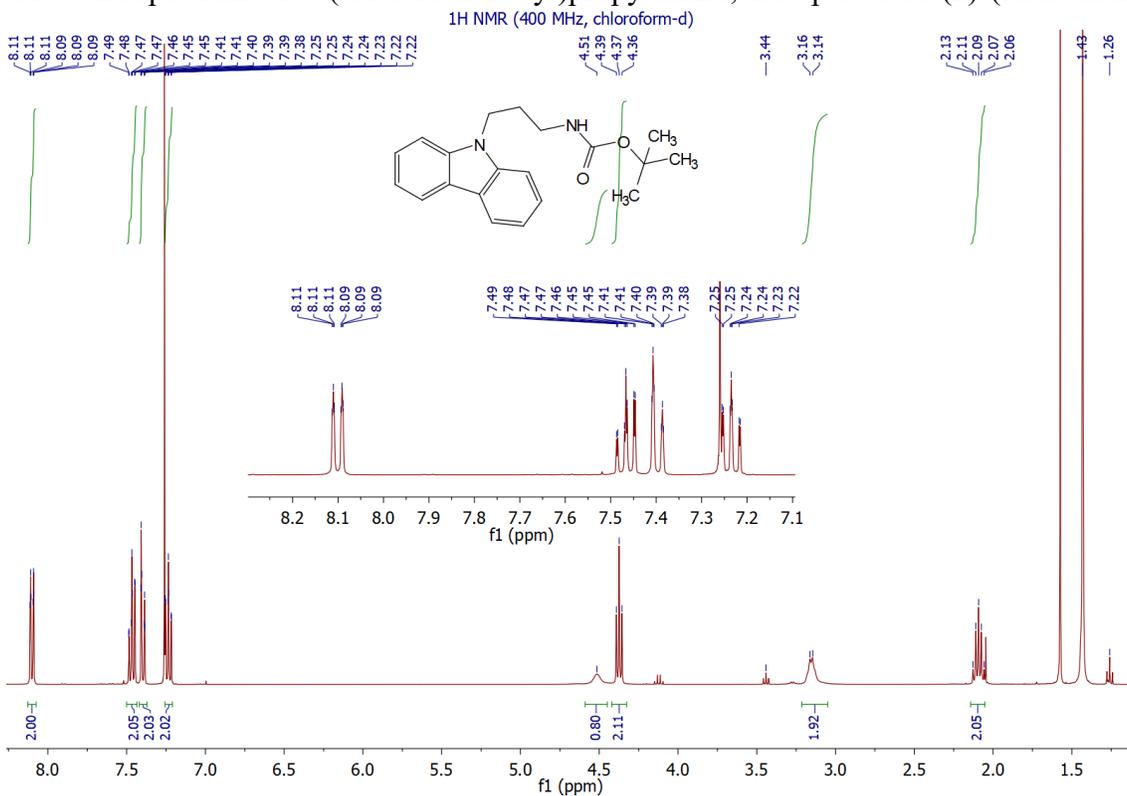



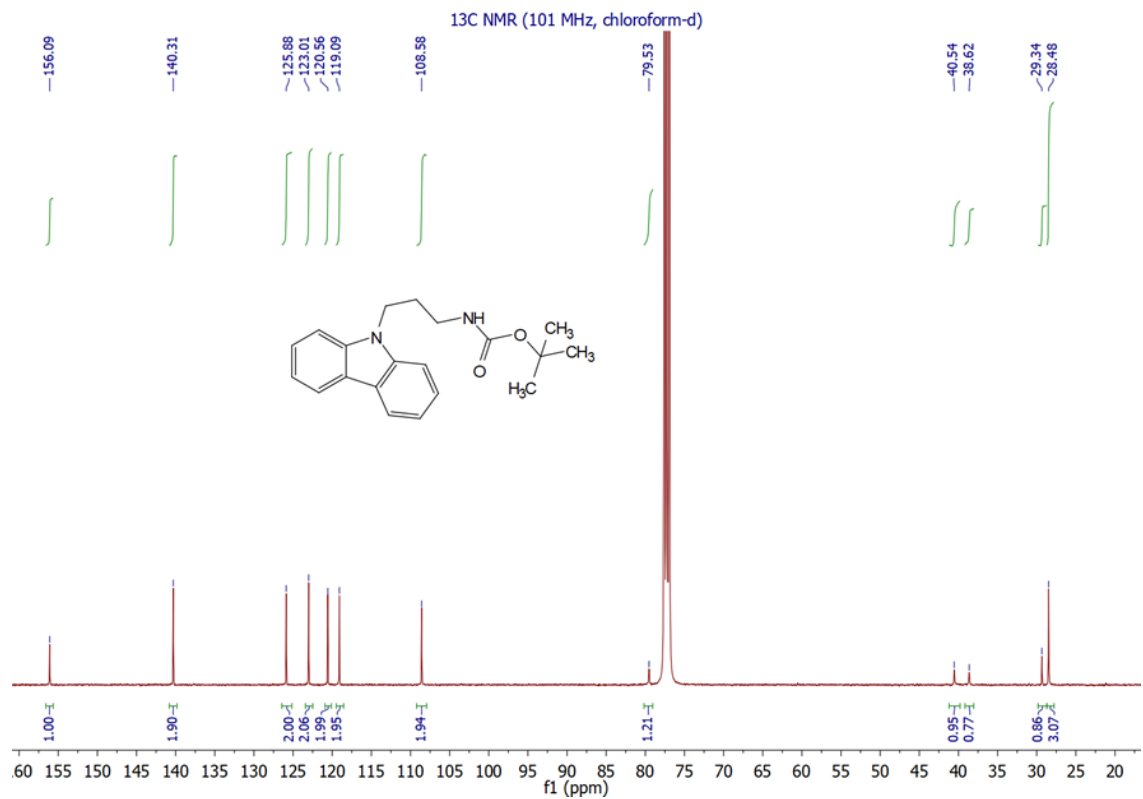

$^{13}$C NMR spectrum of 3-(9H-carbazol-9-yl)propylamine, Boc protected (**2**) (solvent signal: CDCl$_3$ at 77 ppm).

### 3-(9H-carbazol-9-yl)propylammonium iodide (Cz-C$_3$I)

HI (57% in water, unstabilized) was extracted three times with a 9:1 mixture of chloroform and tri-*n*-butyl phosphate to remove impurities. 895 µL (6.78 mmol) of this freshly extracted HI was then added with a micropipette to a solution of **2** (1.00 g, 3.08 mmol) in 20 mL dioxane at 40°C under Ar atmosphere. After stirring for 22 h in the dark, the solvent was partially evaporated, and the remaining solution was added dropwise to 400 mL diethyl ether. The target compound precipitates and was separated after filtering under vacuum and washing with a copious amount of diethyl ether. 0.98 g (91% yield) was obtained as a yellow powder. It was dried under high vacuum before further use. $^1$H NMR (400 MHz, DMSO-$d_6$) δ 8.17 (dt, J = 7.7, 1.0 Hz, 2H), 7.67 (dt, J = 8.2, 0.9 Hz, 2H), 7.61 (s, 3H), 7.48 (ddd, J = 8.3, 7.1, 1.2 Hz, 2H), 7.22 (ddd, J = 7.9, 7.1, 0.9 Hz, 2H), 4.50 (t, J = 6.9 Hz, 2H), 2.87 – 2.77 (m, 2H), 2.05 (dt, J = 14.6, 7.1 Hz, 2H). $^{13}$C NMR (101 MHz, Chloroform-*d*) δ 140.39, 126.35, 122.66, 120.99, 119.52, 37.54, 27.39.



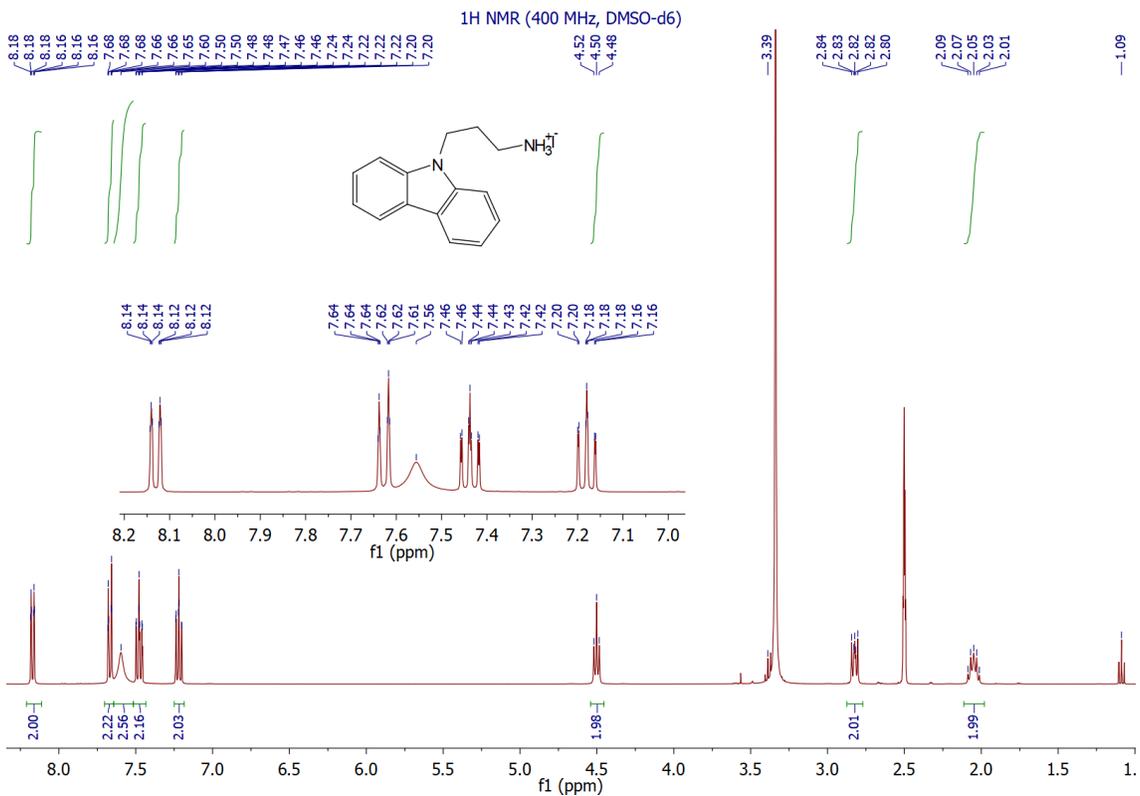

$^1$H NMR spectrum of 3-(9$H$-carbazol-9-yl)propylammonium iodide (Cz-C$_3$I) (residual internal (CD$_3$)(CD$_2$H)SO at 2.50 ppm; water at 3.34 ppm; diethylether at 3.39 ppm and 1.09 ppm).

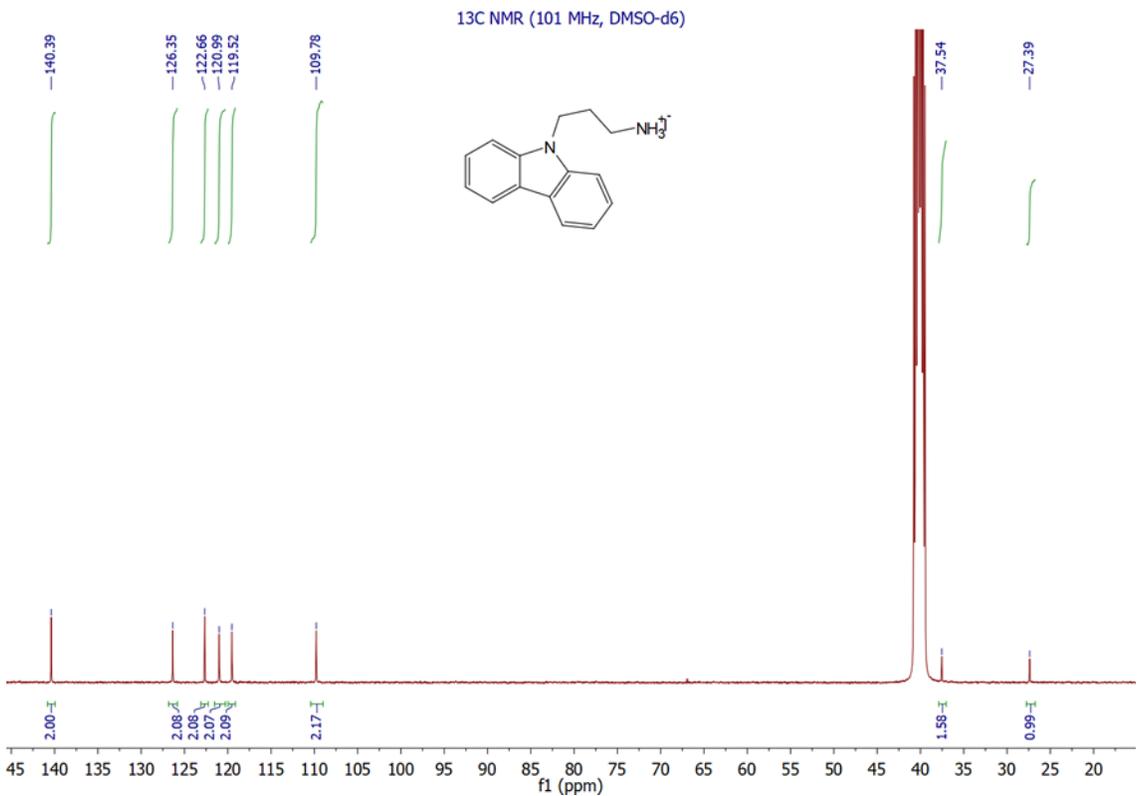

$^{13}$C NMR spectrum of 3-(9$H$-carbazol-9-yl)propylammonium iodide (Cz-C$_3$I) (solvent signal: (CD$_3$)$_2$SO at 40 ppm).



The Cz-C$_4$I and Cz-C$_5$I salts were synthesized according to previous work by some of the authors.[2,3]

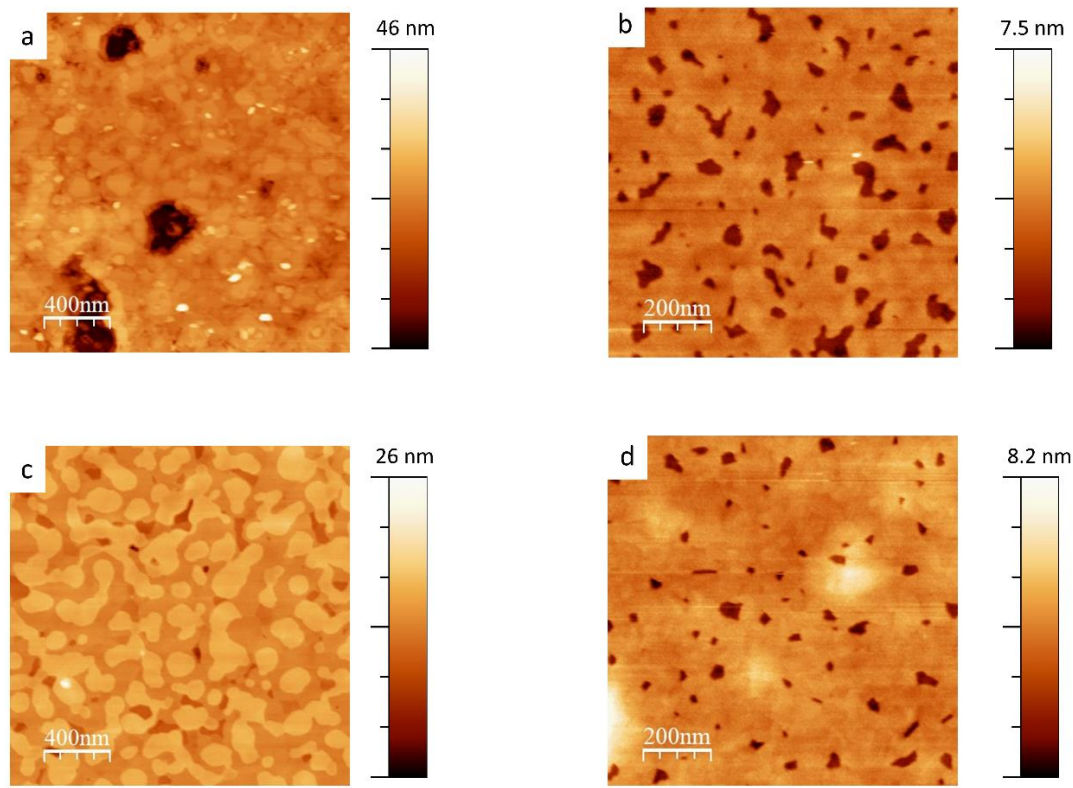

**Figure S1** Atomic force microscopy images of (PEA)$_2$PbI$_4$ (a), (Cz-C$_3$)$_2$PbI$_4$ (b), (Cz-C$_4$)$_2$PbI$_4$ (c) and (Cz-C$_5$)$_2$PbI$_4$ (d). Thicknesses ($w$) are 17 nm, 38 nm, 34 nm and 46 nm, respectively.

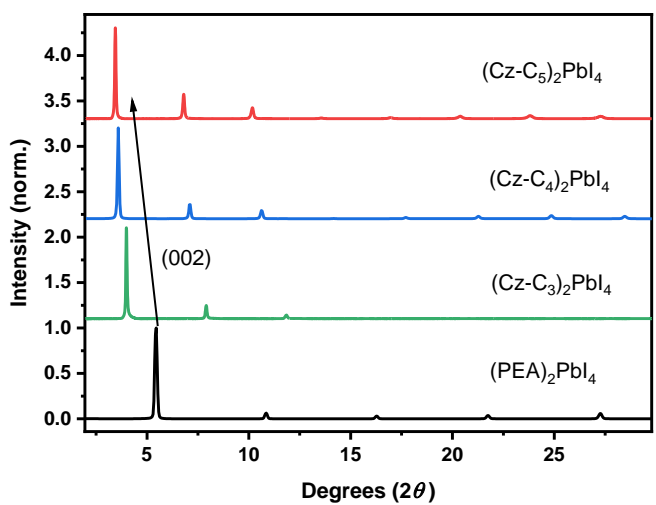



**Figure S2** X-ray diffraction patterns of (PEA)$_2$PbI$_4$, (Cz-C$_3$)$_2$PbI$_4$, (Cz-C$_4$)$_2$PbI$_4$ and (Cz-C$_5$)$_2$PbI$_4$ thin films. The arrow indicates that the position of the (002) reflection shifts to lower diffraction angle (2θ) with increasing alkyl spacer length, indicating an increase in interplanar spacing (d-spacing).

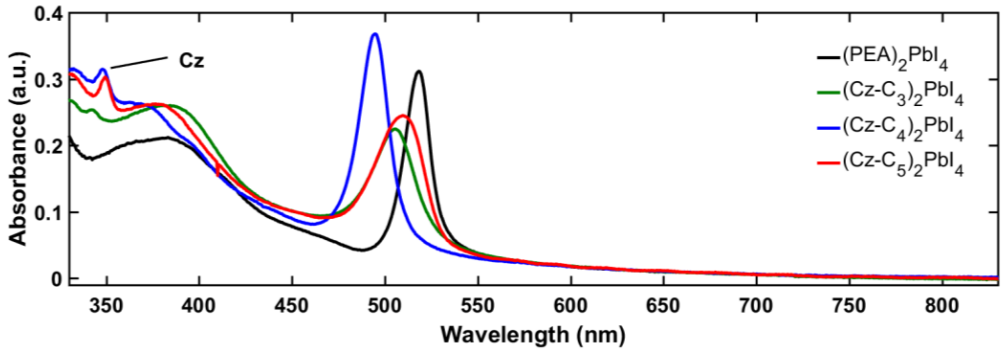

**Figure S3** UV/VIS absorption spectra of (PEA)$_2$PbI$_4$, (Cz-C$_3$)$_2$PbI$_4$, (Cz-C$_4$)$_2$PbI$_4$ and (Cz-C$_5$)$_2$PbI$_4$ thin films. The first excited state of Cz-C$_n$ is indicated with a bar. Note the small spectral shifts.

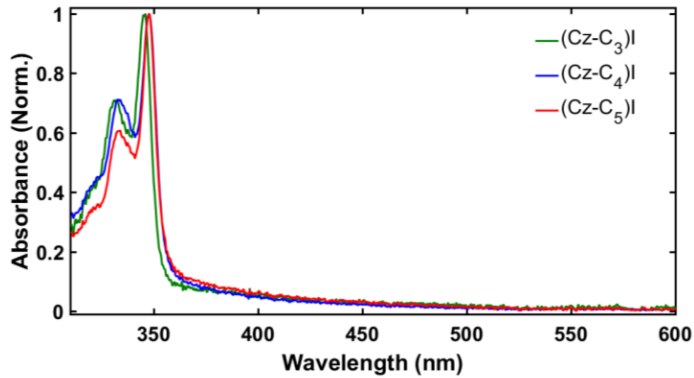

**Figure S4** Normalized UV/VIS Absorption spectra of the carbazole alkyl ammonium salts.

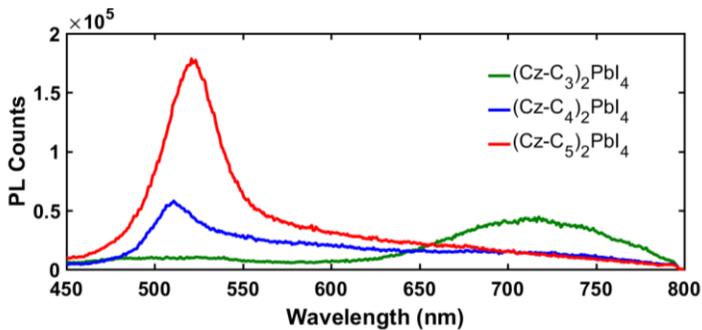

**Figure S5** Photoluminescence spectra of (Cz-C$_3$)$_2$PbI$_4$, (Cz-C$_4$)$_2$PbI$_4$ and (Cz-C$_5$)$_2$PbI$_4$ thin films. λ$_{exc}$ = 400 nm (0.8 mJ/cm$^2$).



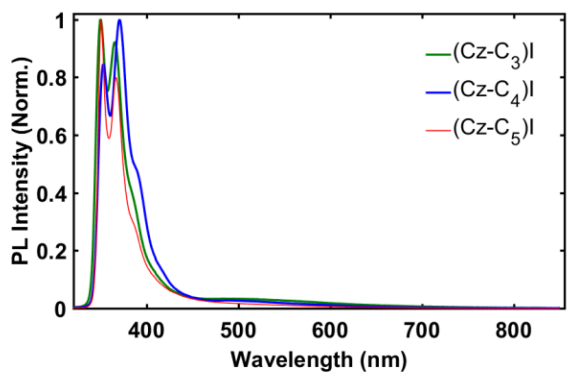

**Figure S6** Normalized photoluminescence spectra of the carbazole alkyl ammonium salts, excited at 300 nm.

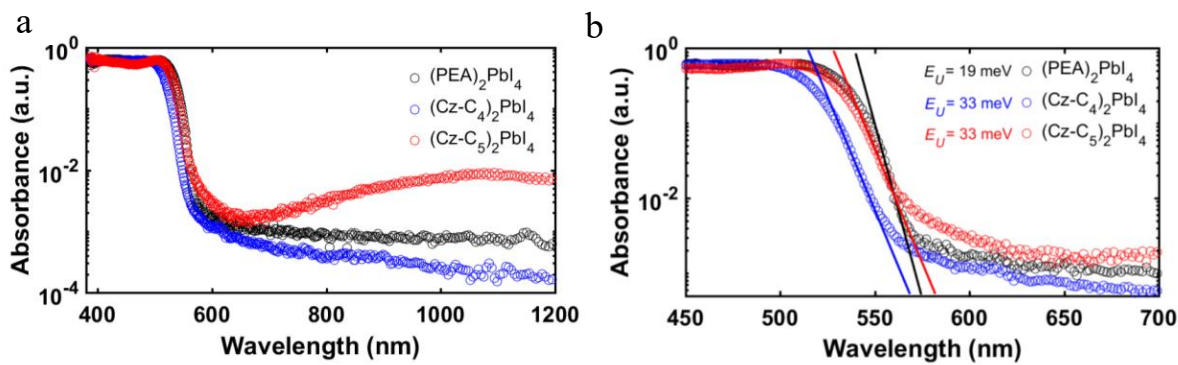

**Figure S7** (a) Normalized photothermal deflection spectra of PEA$_2$PbI$_4$, (Cz-C$_4$)$_2$PbI$_4$ and (Cz-C$_5$)$_2$PbI$_4$ thin films. (b) zoom-in on the sub-gap absorption tail. Urbach energies are extracted from the slope.

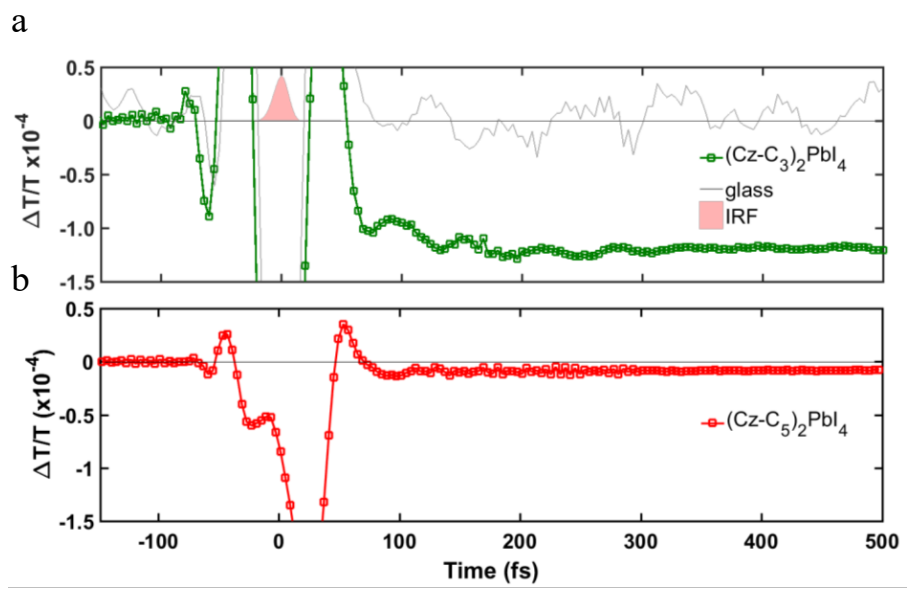

**Figure S8** Femtosecond transient absorption spectroscopy of (Cz-C$_3$)$_2$PbI$_4$ (a) and (Cz-C$_5$)$_2$PbI$_4$ (b) thin films, showing the rise kinetics of the same spectral region (790-820 nm) that was investigated for (Cz-C$_3$)$_2$PbI$_4$ in Figure 2 of the main text. The same pump pulse, also shown in this figure, was used. The



instrument response function (IRF) is determined by the temporal width of this pulse, which is shown in (a).

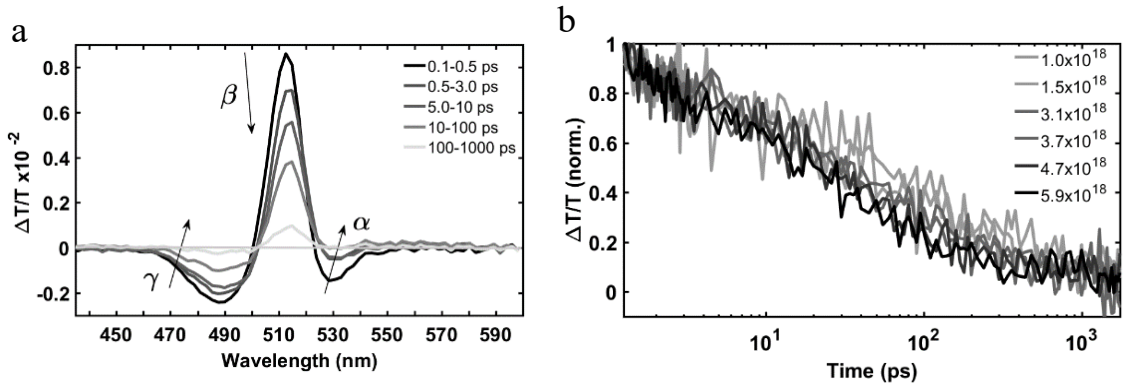

**Figure S9** (a) Transient absorption spectra of a (PEA)$_2$PbI$_4$ thin film excited at 400 nm (5.6 μJ/cm$^2$) integrated over different time regimes. The labels α, β and γ are discussed in the main text. (b) Kinetics of the exciton bleach (510-520 nm, β) of (PEA)$_2$PbI$_4$ thin film excited at 400 nm with different carrier densities (cm$^{-3}$), normalized at 1 ps.

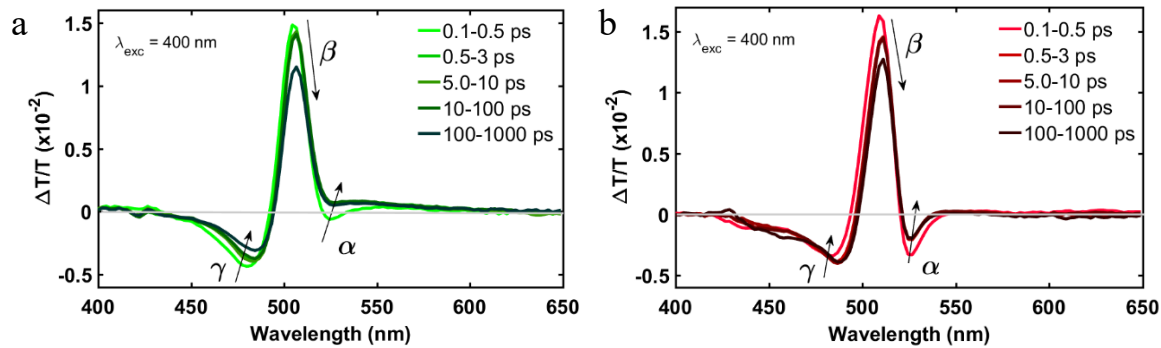

**Figure S10** Transient absorption spectra of (Cz-C$_3$)$_2$PbI$_4$ (a) and (Cz-C$_5$)$_2$PbI$_4$ (b) thin films excited at 400 nm integrated over different time regimes. The fluences (carrier densities) were 9.3 μJ/cm$^2$ (1.6·10$^{18}$ cm$^{-3}$) and 4.2 μJ/cm$^2$ (6.5·10$^{17}$ cm$^{-3}$), respectively. The labels α, β and γ are discussed in the main text.



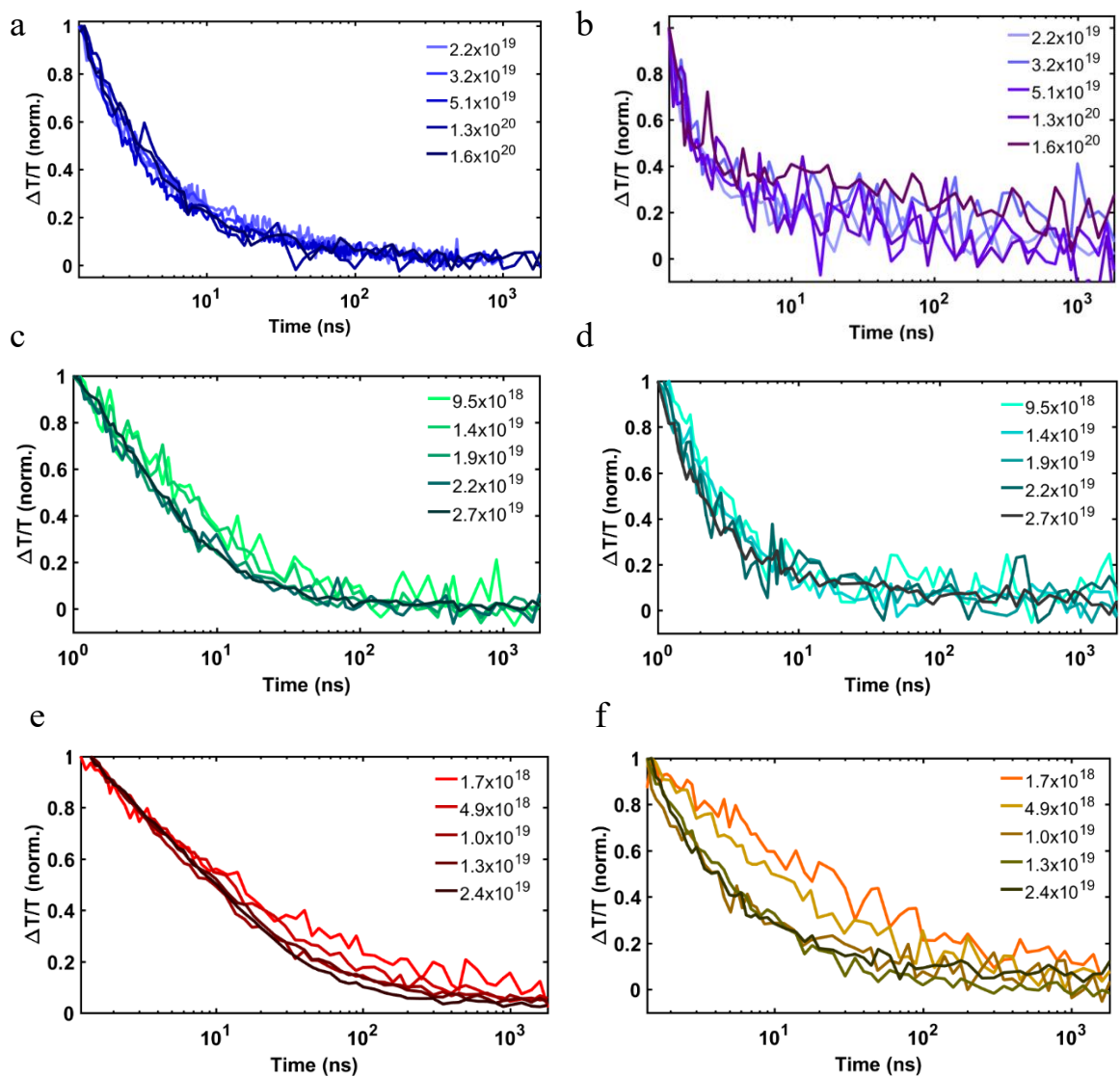

**Figure S11** Normalized nanosecond kinetics of the exciton bleach (β) and photoinduced absorption ($Cz^+$) for $(Cz-C_4)_2PbI_4$ (a,b), $(Cz-C_3)_2PbI_4$ (c,d) and $(Cz-C_5)_2PbI_4$ (e,f) thin films excited at 400 nm with different carrier densities ($cm^{-3}$).



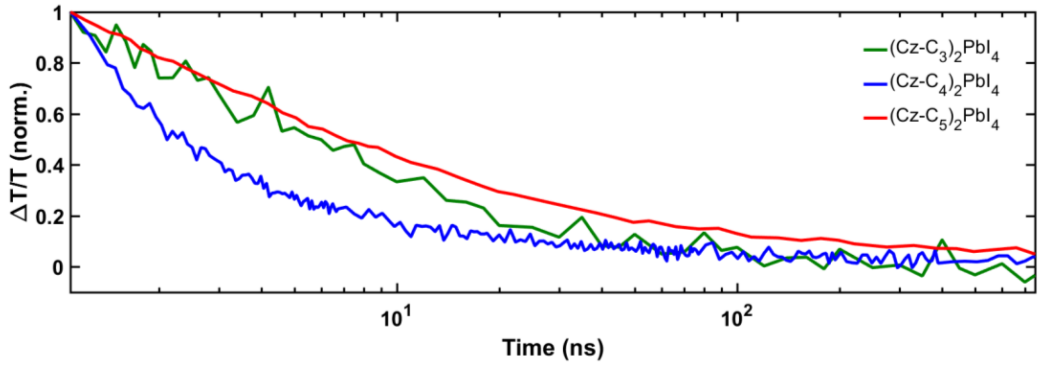

**Figure S12** Normalized nanosecond kinetics of the exciton bleach (β) for (Cz-C$_4$)$_2$PbI$_4$, (Cz-C$_3$)$_2$PbI$_4$ and (Cz-C$_5$)$_2$PbI$_4$ thin films excited at 400 nm (1x10$^{19}$ cm$^{-3}$).

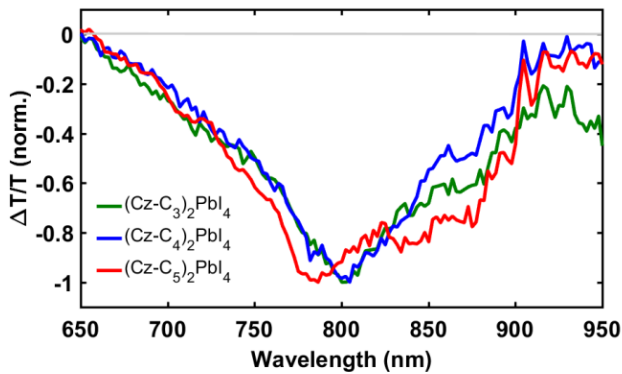

**Figure S13** Normalized transient absorption spectra for (Cz-C$_3$)$_2$PbI$_4$, (Cz-C$_4$)$_2$PbI$_4$ and (Cz-C$_5$)$_2$PbI$_4$ thin films excited at their 1s transition (495 nm for (Cz-C$_4$)$_2$PbI$_4$ and 505 nm for (Cz-C$_3$)$_2$PbI$_4$ and (Cz-C$_5$)$_2$PbI$_4$ with carrier densities of 1x10$^{19}$ cm$^{-3}$ and integrated from 1-3 ps.

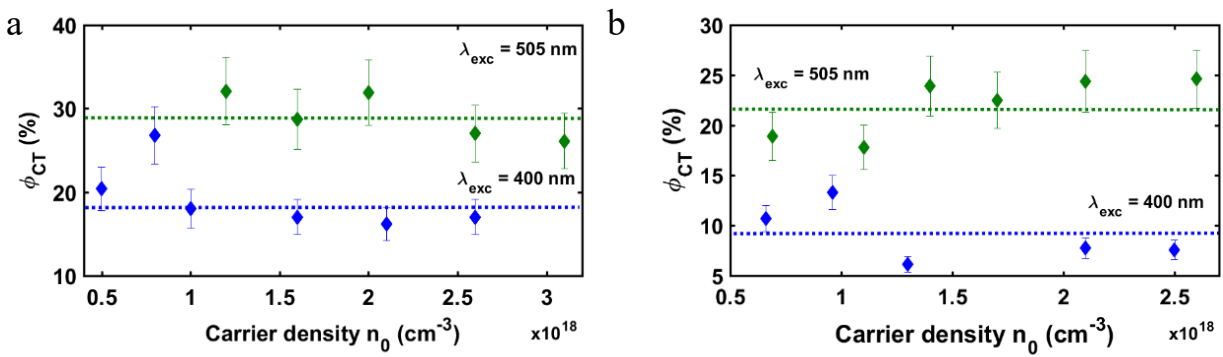

**Figure S14** Charge transfer quantum yields $\phi_{CT}$ for (Cz-C$_3$)$_2$PbI$_4$ and (Cz-C$_5$)$_2$PbI$_4$ thin films excited at 400 nm and their 1s transition (505 nm) with different carrier densities. The dotted lines indicate the average $\phi_{CT}$ values which are used in Fig. 3b.



a

b



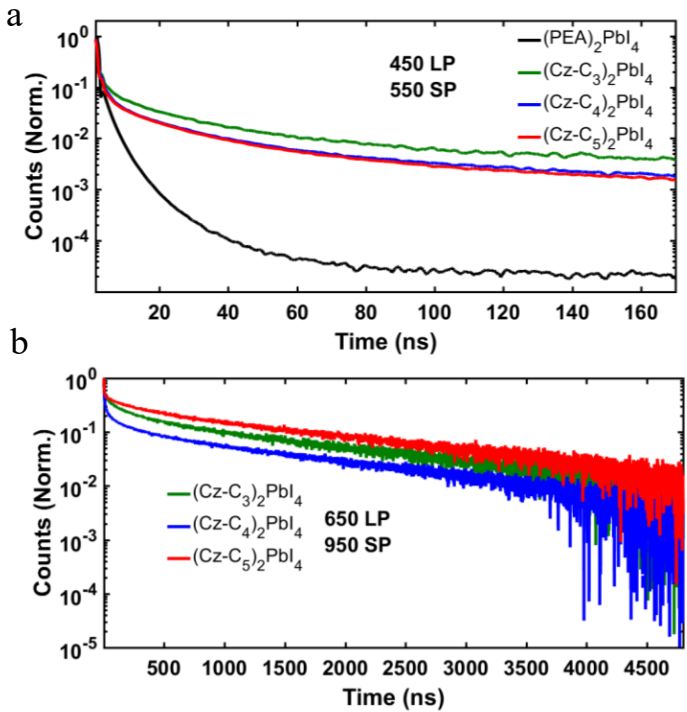

**Figure S15** Normalized time-correlated single photon counting decay curves of (PEA)$_2$PbI$_4$ and (Cz-C$_i$)$_2$PbI$_4$ films photoexcited at 407 nm (0.10 μJ/cm$^2$) using two different sets of filters to selectively detect either the blue (a) or red (b) emission (see Figure S5).

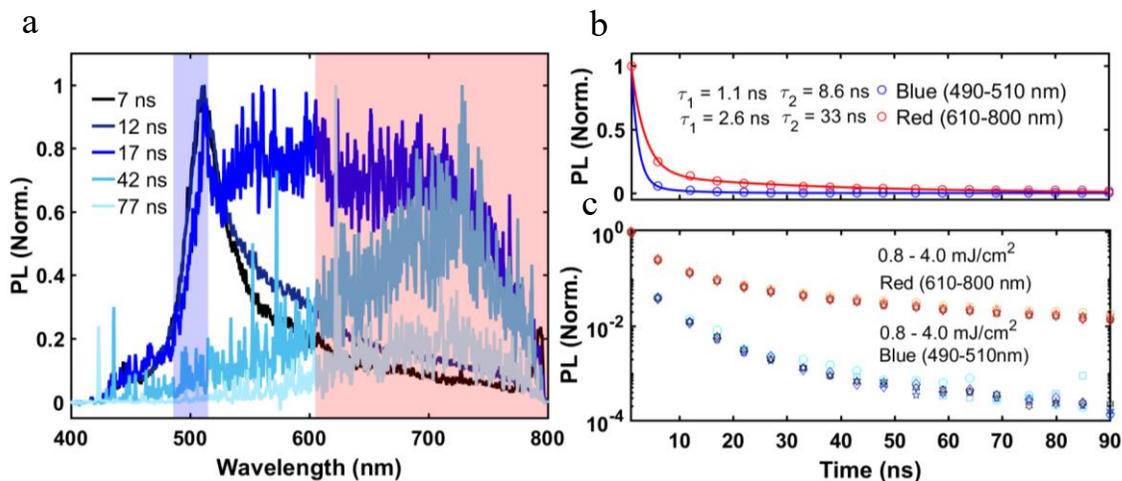

**Figure S16** (a) Normalized time-resolved photoluminescence spectra of a (Cz-C$_4$)$_2$PbI$_4$ thin film photo-excited at 400 nm (0.8 mJ/cm$^2$). (b) Nanosecond photoluminescence kinetics of blue and red spectral regions, which are indicated with the shaded boxes in (a), and their corresponding bi-exponential fits. (c) Fluence dependent photoluminescence kinetics of blue and red spectral regions, 0.8 mJ/cm$^2$, 1.6 mJ/cm$^2$, 2.4 mJ/cm$^2$, 3.2 mJ/cm$^2$, 4.0 mJ/cm$^2$.



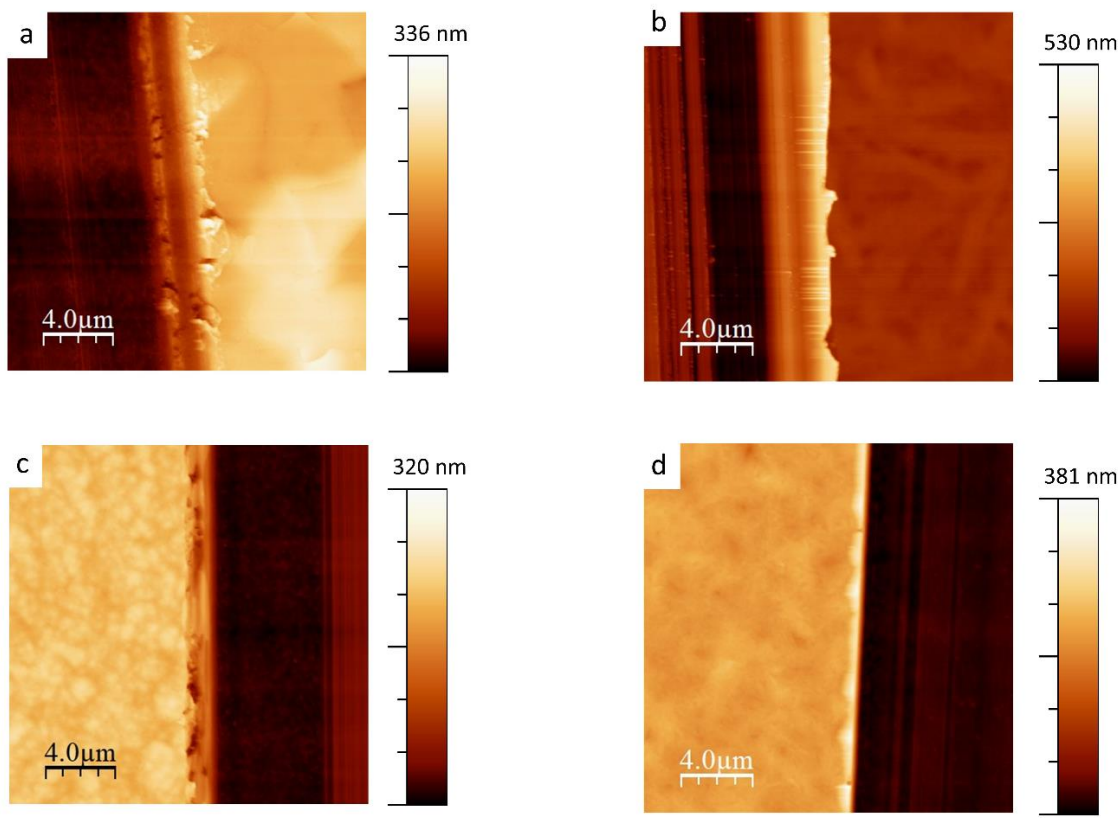

**Figure S17** Atomic Force Microscopy images of $(PEA)_2PbI_4$ (a), $(Cz-C_3)_2PbI_4$ (b) $(Cz-C_4)_2PbI_4$ (c) and $(Cz-C_5)_2PbI_4$ (d) films spin-coated on ITO substrates. Averaged thicknesses of multiple films are provided in Table S2.

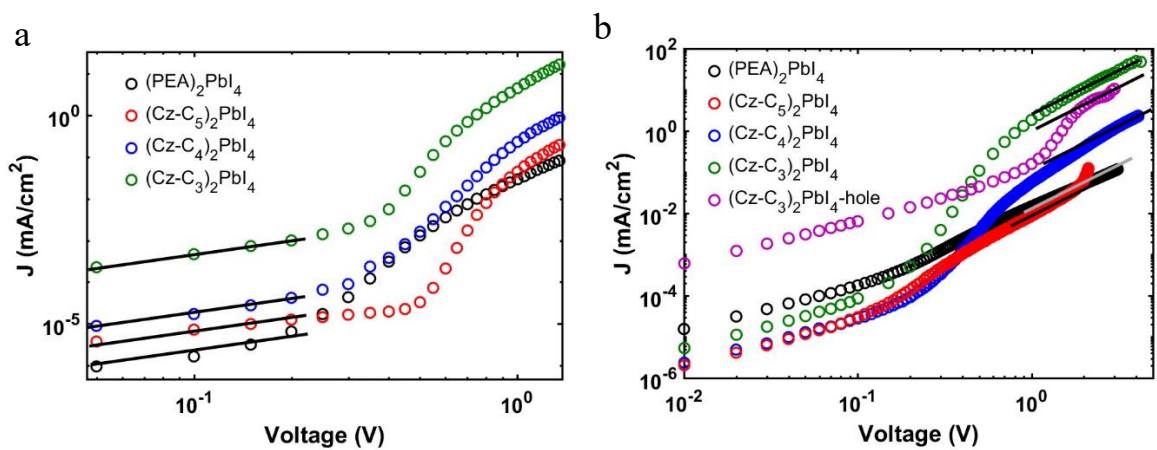

**Figure S18** Representative J-V curves across multiple voltage regimes measured on several 2D perovskite vertical transport devices (shown in Figure 5 of the main text). (a) ambipolar devices used for out-of-plane conductivity ($\sigma_{OOP}$) determination. The black lines indicate the linear (ohmic) regime. (b) electron-selective devices used for out-of-plane mobility ($\mu_{OOP}$) determination. The black lines (grey for



(PEA)$_2$PbI$_4$) indicate the quadratic (SCLC) regime. The J-V curve of the hole-selective device for (Cz-C$_3$)$_2$PbI$_4$ is shown as well.

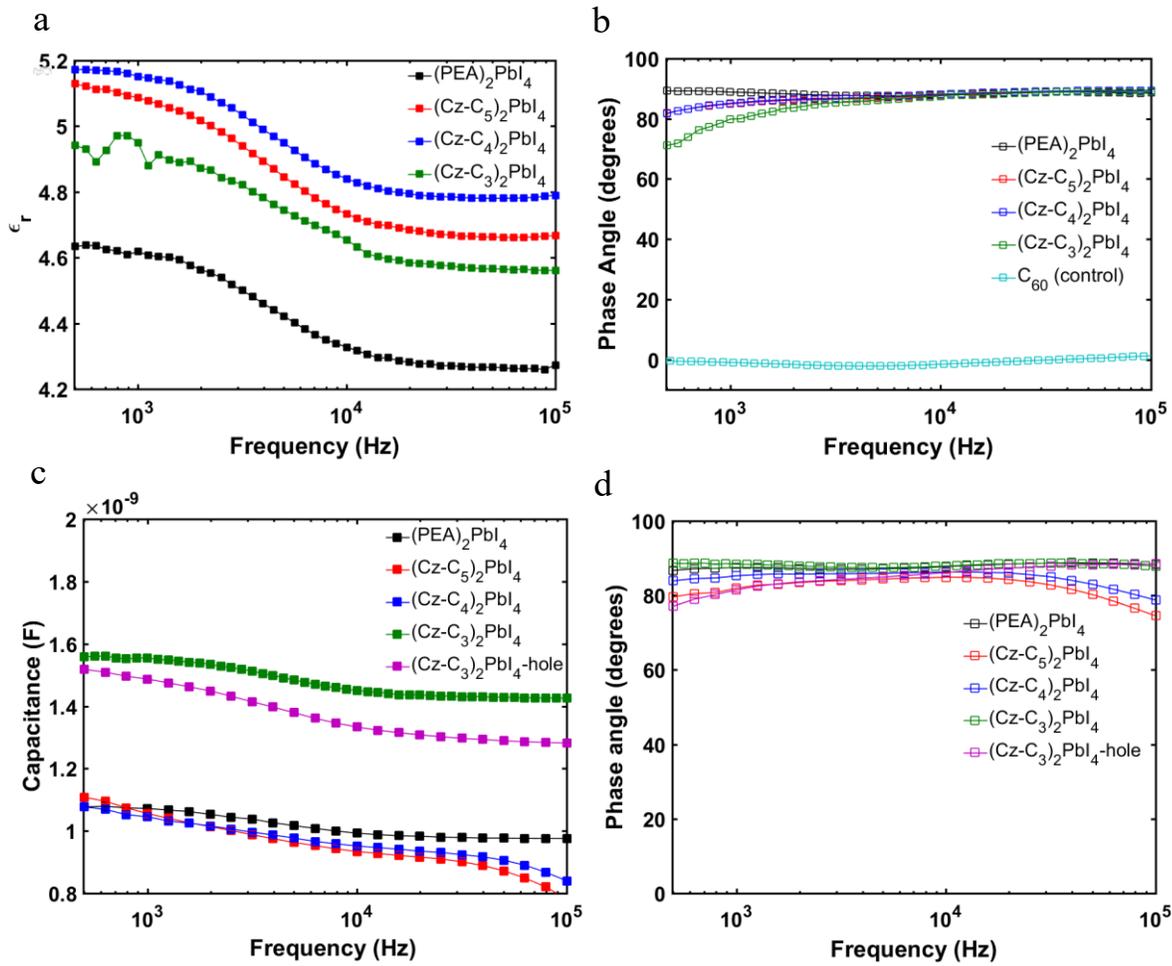

**Figure S19** (a) relative dielectric constant ($\varepsilon_r$) of the 2D perovskite ambipolar devices (Figure 5 main text) by capacitance-frequency (C-F) measurement using the thickness determined from AFM, see Table S2. (b) Phase angle curves of the C-F measurement for the same devices and a Au/C$_{60}$(20 nm)/ITO control device. (c) C-F measurement for electron-only transport devices. Capacitance at 500 Hz is used for out-of-plane mobility determination. (d) Phase angle curves of the C-F measurements for the same devices.

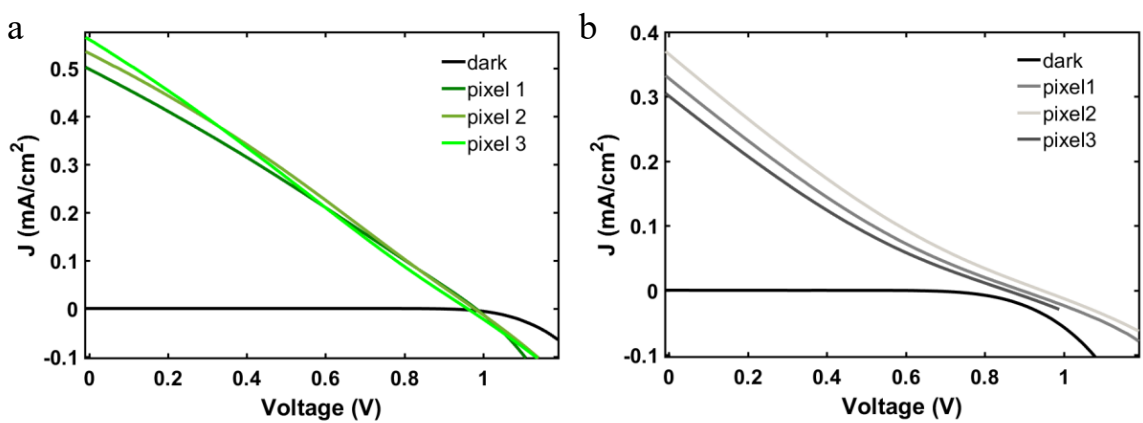



**Figure S20** J-V curves for a photovoltaic device of (Cz-C$_3$)$_2$PbI$_4$ (a) and (PEA)$_2$PbI$_4$ (b). Corresponding parameters are provided in Table S1.

**Table S1. Extracted photovoltaic parameters of J-V curves shown in Figure S20.**

|  | V$_{OC}$ (V) | J$_{SC}$ (mA/cm$^2$) | FF (%) | PCE (%) |
|---|---|---|---|---|
| Pixel 1 (Cz-C$_3$)$_2$PbI$_4$ | 0.980 | 0.502 | 26.8 | 0.132 |
| Pixel 2 (Cz-C$_3$)$_2$PbI$_4$ | 0.978 | 0.535 | 27.3 | 0.143 |
| Pixel 3 (Cz-C$_3$)$_2$PbI$_4$ | 0.959 | 0.564 | 25.3 | 0.137 |
| Pixel 1 (PEA)$_2$PbI$_4$ | 0.888 | 0.333 | 19.6 | 0.058 |
| Pixel 2 (PEA)$_2$PbI$_4$ | 0.943 | 0.370 | 19.8 | 0.069 |
| Pixel 3 (PEA)$_2$PbI$_4$ | 0.858 | 0.305 | 19.5 | 0.051 |

**S1 Determination of charge transfer quantum yield and experimental error**

The charge transfer quantum yield, denoted as $\varphi_{CT}$, is defined as the ratio of Cz$^{+\cdot}$ molecules generated per unit volume to the number of absorbed photons per unit volume, $n_0$:

$$\varphi_{CT} (\%) = \frac{Cz^{+\cdot}}{n_0} \times 100 \tag{S1}$$

$Cz^{+\cdot}$ (cm$^{-3}$) is related to the TA signal at a certain probe wavelength and time delay, $\frac{\Delta T}{T}$, the absorption coefficient at the same probe wavelength, $\sigma$ (cm$^2$), and the film thickness, $w$ (cm), through

$$\frac{\Delta T}{T} = -w \times \sigma \times Cz^{+\cdot} \tag{S2}$$

We use AFM to measure $w$ (Figure S1, SI) and we take $\sigma$ from ref. [4] To calculate $n_0$ and to correct for the different sizes in pump and probe pulses, we follow the procedure of ref. [5] which takes into account the Gaussian shapes of the pulses. The improved $n_0$ is calculated through equation S3:

$$n_0 = \frac{F_{corr}(1-10^{-A})}{wE_{photon}} \tag{S3}$$

where F is the pump fluence, A is the absorbance at the pump wavelength, w is the film thickness (this expression is valid since the absorption depth >> film thickness) and E$_{photon}$ is the photon energy of the pump. The estimated amount of Cz$^{+\cdot}$ absorption determined from the probe beam absorption assumes the same vertically uniform laser profile. However, the pump and probe beam spot sizes are significantly different in our experiments. As mentioned in the main text, we use the procedure in ref.x[5] to correct for this. The pump intensity in the region of pump-probe beam overlap, I$_{pump}$(r$_{probe}$), is calculated using equation S2

$$I_{pump}(r_{probe}) = I_{tot}\left[1 - \exp\left(\frac{-2r_{probe}^2}{r_{pump}^2}\right)\right] \tag{S4}$$

where I$_{tot}$ is the total intensity of the pump beam on the film, r$_{probe}$ is the probe beam spot size and r$_{pump}$ is the pump beam spot size. The latter two are extracted by taking 1/e$^2$ radius of a Gaussian fit. Using this corrected pump intensity, F is computed according to equation S3 and substituted in S1.



$$F_{corr} = \frac{I_{pump}(r_{probe})}{(\pi \times r_{probe}^2) \times f} \tag{S5}$$

in which f is the repetition rate of the laser.

The total experimental errors are mainly caused by errors in $I_{tot}$, $r_{probe}$ and $r_{pump}$. We use error propagation schemes based on relative errors of 6%, 3% and 5% for $I_{tot}$, $r_{probe}$ and $r_{pump}$, respectively, to yield the error of 13% in $I_{pump}(r_{probe})$. The absolute charge transfer quantum yield values are multiplied by this relative error to yield the absolute experimental error.

**S2 Determination of out-of-plane conductivity and charge carrier mobility and experimental error**

Out-of-plane conductivities ($\sigma_{OOP}$) were determined from the J-V curves in the ohmic regime, where there is a linear relationship between current density and voltage. The $\sigma_{OOP}$ is determined by multiplying the slope, $a = \left(\frac{dJ}{dV}\right)$, with the perovskite film thickness L, as determined by AFM. The error in $\sigma_{OOP}$ has contributions from both as expressed in Equation S6:

$$\left(\frac{\Delta\sigma}{\sigma}\right)^2 = \left(\frac{\Delta a}{a}\right)^2 + \left(\frac{\Delta L}{L}\right)^2 \tag{S6}$$

Out-of-plane mobilities ($\mu_{OOP}$) were determined from the J-V curves in the space-charge limited current (SCLC) regime, where there is a quadratic relationship between the current density and voltage. This is expressed in the Mott-Gurney equation as:

$$J = \frac{9}{8} * \varepsilon * \mu * \frac{V^2}{L^3} \tag{S7}$$

where $J$ is the current density, $\varepsilon$ is the dielectric constant, $\mu$ is the mobility, $L$ is the perovskite film thickness.

$J = I/A$ and $C_{perovskite} = \frac{\varepsilon A}{L}$, where A is the area of the device pixel. As we are using the same device pixel for the I-V and capacitance measurements, the area term A may be cancelled by the following transformation:

$$\frac{I}{A} = \frac{9}{8} * \left(\frac{C_{perovskite} * L}{A}\right) * \mu * \frac{V^2}{L^3} \tag{S8}$$

$$I = \frac{9}{8} * C_{perovskite} * \mu * L^{-2} * V^2 \tag{S9}$$

And we finally obtain the equation for the extraction of mobility after some further rearranging:

$$\sqrt{I} = \left(\frac{9}{8} * C_{perovskite} * \mu * L^{-2}\right)^{1/2} * V \tag{S10}$$

$$\left(\frac{d\sqrt{I}}{dV}\right)^2 = \frac{9}{8} * C_{perovskite} * \mu * L^{-2} \tag{S11}$$

$$\mu = \frac{8}{9} * \left(\frac{d\sqrt{I}}{dV}\right)^2 * \frac{L^2}{C_{perovskite}} \tag{S12}$$



Errors in the extracted mobility have three contributions: 1) SCLC regime slope $k = \frac{d\sqrt{I}}{dV}$, for which the error is the standard deviation of the linear regression ($\Delta k$), 2) the error in perovskite film thickness, $\Delta L$, and 3) the error in capacitance, $\Delta C_{perovskite}$, measured with impedance spectroscopy. The standard deviation of mobility ($\Delta \mu$) is then given by the following equation:

$$\left(\frac{\Delta \mu}{\mu}\right)^2 = \left(\frac{2\Delta k}{k}\right)^2 + \left(\frac{2\Delta L}{L}\right)^2 + \left(\frac{-\Delta C_{perovskite}}{C_{perovskite}}\right)^2 \quad \text{(S13)}$$

Finally, there is an error associated with mobilities determined from different $N$ devices. The average standard deviation of the mobility $\langle \Delta \mu \rangle$ is determined by the following equation:

$$\langle \Delta \mu \rangle = \sqrt{\frac{1}{N} * \Sigma_i(\mu_i^2 + \Delta \mu_i^2) - \left(\frac{1}{N} * \Sigma_i \mu_i\right)^2} \quad \text{(S14)}$$

All relevant parameters and corresponding errors are provided in Table S2 and S3 for out-of-plane conductivity and mobility determination, respectively.

**Table S2. Parameters with corresponding errors used for out-of-plane conductivity determination.**

| | $a$ (mA/V·cm$^2$) | $L$ (nm) | $\sigma_{OOP}$ (S/cm) |
|---|---|---|---|
| (PEA)$_2$PbI$_4$ | 2.03(±0.25)x10$^{-5}$ | 140±14 | 2.84(±0.45)x10$^{-13}$ |
| (Cz-C$_5$)$_2$PbI$_4$ | 4.48(±0.28)x10$^{-5}$ | 208±21 | 9.32(±1.1)x10$^{-13}$ |
| (Cz-C$_4$)$_2$PbI$_4$ | 2.42(±0.31)x10$^{-4}$ | 159±16 | 3.84(±0.63)x10$^{-12}$ |
| (Cz-C$_3$)$_2$PbI$_4$ | 6.33(±0.46)x10$^{-3}$ | 87±9 | 5.51(±0.68)x10$^{-11}$ |

**Table S3. Parameters with corresponding errors used for out-of-plane mobility determination.**

| | $k$ ($\sqrt{A}/V$) | $C$ (F) x10$^{-9}$ | $L$ (nm) | $\mu_{OOP}$ (cm$^2$/V·s) |
|---|---|---|---|---|
| (PEA)$_2$PbI$_4$ electron | 7.54(±0.05)x10$^{-4}$ | 1.077±0.1 | 166±15 | 1.31(±0.48)x10$^{-7}$ |
| (Cz-C$_5$)$_2$PbI$_4$ electron | 1.09(±0.02)x10$^{-3}$ | 1.108±0.1 | 193±8 | 3.55(±0.45)x10$^{-7}$ |
| (Cz-C$_4$)$_2$PbI$_4$ electron | 3.89(±0.02)x10$^{-3}$ | 1.077±0.1 | 206±10 | 5.87(±3.62)x10$^{-6}$ |
| (Cz-C$_3$)$_2$PbI$_4$ electron | 1.18(±0.05)x10$^{-2}$ | 1.597±0.1 | 117±14 | 1.06(±0.30)x10$^{-5}$ |
| (Cz-C$_3$)$_2$PbI$_4$ hole | 6.70(±0.51)x10$^{-3}$ | 1.440±0.1 | 117±14 | 3.85(±1.28)x10$^{-6}$ |